\documentstyle[preprint,array,aps,prc,rotate]{revtex}
\begin{document}
\input{psfig}
\input{epsf}
\def\Im{\mbox{\sl Im\ }}
\def\pd{\partial}
\def\oln{\overline}
\def\olft{\overleftarrow}
\def\ds{\displaystyle}
\def\bgreek#1{\mbox{\boldmath $#1$ \unboldmath}}
\def\sla#1{\slash \hspace{-2.5mm} #1}
\newcommand{\bra}{\langle}
\newcommand{\ket}{\rangle}
\newcommand{\vep}{\varepsilon}
\newcommand{\met}{{\mbox{\scriptsize met}}}
\newcommand{\lab}{{\mbox{\scriptsize lab}}}
\newcommand{\cm}{{\mbox{\scriptsize cm}}}
\newcommand{\mcal}{\mathcal}
\newcommand{\Del}{$\Delta$}
\newcommand{\g}{{\rm g}}
\long\def\Omit#1{}
\long\def\omit#1{\small #1}
\def\beq{\begin{equation}}
\def\eeq{\end{equation} }
\def\bea{\begin{eqnarray}}
\def\eea{\end{eqnarray}}
\def\eqref#1{Eq.~(\ref{eq:#1})}
\def\eqlab#1{\label{eq:#1}}
\def\figref#1{Fig.~(\ref{fig:#1})}
\def\figlab#1{\label{fig:#1}}
\def\tabref#1{Table \ref{tab:#1}}
\def\tablab#1{\label{tab:#1}}
\def\secref#1{Section~\ref{sec:#1}}
\def\seclab#1{\label{sec:#1}}
\def\VYP#1#2#3{{\bf #1}, #3 (#2)}  % Volume, page (Year)
\def\NP#1#2#3{Nucl.~Phys.~\VYP{#1}{#2}{#3}}
\def\NPA#1#2#3{Nucl.~Phys.~A~\VYP{#1}{#2}{#3}}
\def\NPB#1#2#3{Nucl.~Phys.~B~\VYP{#1}{#2}{#3}}
\def\PL#1#2#3{Phys.~Lett.~\VYP{#1}{#2}{#3}}
\def\PLB#1#2#3{Phys.~Lett.~B~\VYP{#1}{#2}{#3}}
\def\PR#1#2#3{Phys.~Rev.~\VYP{#1}{#2}{#3}}
\def\PRC#1#2#3{Phys.~Rev.~C~\VYP{#1}{#2}{#3}}
\def\PRD#1#2#3{Phys.~Rev.~D~\VYP{#1}{#2}{#3}}
\def\PRL#1#2#3{Phys.~Rev.~Lett.~\VYP{#1}{#2}{#3}}
\def\FBS#1#2#3{Few-Body~Sys.~\VYP{#1}{#2}{#3}}
\def\AP#1#2#3{Ann.~of Phys.~\VYP{#1}{#2}{#3}}
\def\ZP#1#2#3{Z.\ Phys.\  \VYP{#1}{#2}{#3}}
\def\ZPA#1#2#3{Z.\ Phys.\ A\VYP{#1}{#2}{#3}}
\def\half{\mbox{\small{$\frac{1}{2}$}}}
\def\quarter{\mbox{\small{$\frac{1}{4}$}}}
\def\nn{\nonumber}
\newlength{\PicSize}
\newlength{\FormulaWidth}
\newlength{\DiagramWidth}
\newcommand{\vslash}[1]{#1 \hspace{-0.5 em} /}

%%%%%%%%%%%%%%%%%%%%%%%%%%%%%%%%%%%%%%%%%%%%%%%%%%%%%%%%%%%%%%%%%%%%%%%%%%
%%     TO REMOVE COMMENTS MARGIN, DELETE COMMANDS \bel,\abo and \her    %%
%%%%%%%%%%%%%%%%%%%%%%%%%%%%%%%%%%%%%%%%%%%%%%%%%%%%%%%%%%%%%%%%%%%%%%%%%%

\def\olaf{\marginpar{Mod-Olaf}}
\def\her{\marginpar{$\Longleftarrow$}}
\def\bel{\marginpar{$\Downarrow$}}
\def\abo{\marginpar{$\Uparrow$}}

%\pagestyle{headings}
%%\pagestyle{myheadings}
%%\markright{\today \hfill
%%     {\it Compton-full, version-x3n}
%%      \hspace{1cm}}

%\oddsidemargin +0.0mm
%\evensidemargin +0.0mm
%\topmargin -13.0mm
%%\voffset 1.5cm  % print at top of page
%%\textheight 24cm
%\textwidth 17cm
%\oddsidemargin .5cm
%\evensidemargin .5cm

%%%%%%%%       {\bf DRAFT! }
%%%%%%%%        \preprint{HEP/123-qed}
%\tighten

\title{Compton scattering on the nucleon at intermediate energies and polarizabilities
in a microscopic model}

\author{S. Kondratyuk}
%\homepage{http://www.triumf.NL/~kondratyuk}
\address{%
%\affiliation{%
TRIUMF, 4004 Westbrook Mall, Vancouver, British Columbia, Canada V6T 2A3}

\author{O.\ Scholten}
%\email{Scholten@kvi.nl}
%\homepage{http://www.KVI.NL/~Scholten}
\address{%
%\affiliation{%
Kernfysisch Versneller Instituut, University of Groningen, 9747 AA
Groningen, The~Netherlands}

\date{\today}
\maketitle

\begin{abstract}

A microscopic calculation of Compton scattering on the
nucleon is presented which encompasses the lowest energies -- 
yielding nucleon
polarizabilities -- and extends to energies of the order of 600 MeV. 
We have used the covariant ``Dressed K-Matrix Model" obeying the symmetry
properties which are appropriate in the different energy regimes.
In particular, crossing symmetry, gauge invariance and unitarity are
satisfied. The 
extent of violation of analyticity (causality) is used as an expansion 
parameter.

\end{abstract}


%\maketitle

%%%%%%%       \tableofcontents

\section{Introduction}

In this paper we develop a relativistic model suitable for a quantitative
description of real Compton ($\gamma N$) scattering at both low and 
intermediate energies (up to the second resonance region).
To achieve this, the model should obey various constraints that are
important in the different energy regimes. At the lowest energies, gauge
invariance, CPT invariance and crossing symmetry are important for the
model to obey low-energy theorems \cite{Low54}. At energies near the
pion production threshold, unitarity and analyticity put strong
constraints on the amplitude \cite{Pfe74,Ber93,Lvo97,Hun97,Bab98,Dre00}.
Since a wide energy span needs to be described, it is most efficient to 
use a relativistic approach.

An obvious starting point may seem to be the Bethe-Salpeter equation
\cite{Sal51} or a 3-dimensional reduction thereof. They have been used
to accurately describe pion-nucleon ($\pi N$) scattering
\cite{Lah99,Pea91,Gro93,Pas98}. However, with the usual choice of the
kernel -- consisting of tree-level diagrams only -- crossing symmetry is
violated. For this reason we have developed an alternative approach,
called the ``Dressed K-matrix Model"\cite{Kon99,Kon00a,Kon01,Kon00b}.
Since this model is based on a K-matrix formalism, unitarity (in the
coupled channel space) and crossing symmetry are easily implemented. The
kernel is formulated with dressed vertices and propagators such that
certain analyticity constraints for the amplitude are incorporated 
at the level of one-particle reducible diagrams. 
Singularities of the
regularization form factor are chosen far away from the kinematic
regime of interest, and thus we ignore the small violations of analyticity
due to the regularization.  
Also, analyticity is not explicitly
incorporated for one-particle irreducible diagrams. 
We argue in \secref{analyt} that
the level of
violation of analyticity can be regarded as an expansion parameter in
the present model. Gauge invariance is exact through the introduction of
contact terms obtained by minimal substitution.

In \secref{model} we present a short recap of the essential ingredients
of the Dressed K-matrix Model. The parameters in the model Lagrangian
are fixed to a large extent from pion-nucleon scattering and pion
photoproduction as presented in \secref{pi-N}. The results for Compton
scattering and nucleon polarizabilities are given in \secref{ComPol}.

\section{The Dressed K-matrix Model} \seclab{model}

Various ingredients of the Dressed K-Matrix Model have been described in
Refs.~\cite{Kon99,Kon00a,Kon01,Kon00b}. Therefore, here we will present only
the essential motivation and arguments behind the formulation of the model.
Throughout we will use a
manifestly relativistic covariant formulation using the notation of
Ref.~\cite{Bjo64}.

Since we want our model to satisfy the important constraint of crossing
symmetry, we embed it in the K-matrix approach. In Appendix A we show
that any model cast in the K-matrix formalism obeys crossing
symmetry if the kernel $K$ itself is crossing symmetric. For a hermitian
kernel, unitarity is also satisfied, even in a coupled $\pi$-N and
$\gamma$-N channel space. 
The problem is thus now reduced to
constructing a suitable kernel $K$ such that also causality -- or analyticity of
the amplitude -- is satisfied. We will write our kernel as a sum of
tree-level diagrams and contact terms (for gauge invariance), as is done
usually. If this were all, the amplitude would not be an
analytic function 
(the real and imaginary parts of the loop corrections 
would not be related to each other through
dispersion relations), which implies that causality would be violated, 
as is the case in
traditional K-matrix models\cite{Gou94,Sch96,Feu98,Kor98,Feu99}.
The essence of our approach lies in the use of 
{\em dressed} vertices and propagators in the kernel $K$, where
the diagrams selected for the dressing
are chosen such that certain analyticity constraints are
implemented in the calculation of the T-matrix.
Effects of the dressing are expressed
in terms of purely {\em real} form factors and self-energy functions.
Our approximation is in the
extent to which analyticity of the amplitude for the process is
satisfied.

The dressing in our formalism is based on including only those diagrams which
are necessary for satisfying analyticity. As discussed in the next section and
more extensively in Refs.~\cite{Kon99,Kon00a}, this dressing corresponds to
loop corrections to vertices and propagators, where every vertex in a loop
correction is in turn corrected by similar loops to satisfy the constraints also
in the strong interaction regime. Since this type of dressing
cannot be captured as a geometric series in the coupling constant, the summation
is done via a numerical iterative expansion. 

\Omit{
In addition to obeying unitarity and crossing symmetry,
another advantage of using the K-matrix formalism is that it allows us
to formulate the dressing equations in terms of propagators and only
{\em half}-off-shell (not full-off-shell) vertices. This greatly
simplifies our approach.} 
In the following we first present the essential
ingredients of the dressing procedure for the $\pi N N$ and $\gamma N N$
vertices and the nucleon self-energy and only thereafter show the full
structure of the K-matrix kernel.

\subsection{The dressing procedure \label{sec:III}}

\subsubsection{The $\pi$NN vertex}

The objective of dressing the vertices and propagators in the present 
approach is solely to improve on the analytic properties of the 
amplitude. The imaginary parts of the amplitude are generated through 
the K-matrix formalism (as imposed by unitarity) and
correspond to 
cut loop corrections where the intermediate particles are taken on 
their mass shell. The real parts have to follow from 
applying dispersion relations to the imaginary parts. We incorporate these real
parts as real vertex and self-energy functions.
Investigating 
this in detail (for a more extensive discussion we refer to 
~\cite{Kon99,Kon00a}) shows that the dressing can be 
formulated in terms of coupled equations, schematically shown in
\figref{piNN}, which generate multiple overlapping loop corrections.
The 
coupled nature of the equations is necessary to 
obey simultaneously unitarity and analyticity.

\Omit{
We maintain the full Lorentz structure of the
nucleon propagator and $\pi N N$ vertices in the dressing procedure.
In particular, the
general half-off-shell $\pi$NN vertex is written as%\cite{Kaz59}
\begin{equation}
\Gamma_{\alpha}(p)=
\tau_{\alpha}\,\gamma^5 \Big[ G_{ps}(p^2) + \Lambda_{+}(p)\,
G_{pv}(p^2)\Big],
\eqlab{pi-vert}
\end{equation}
where the final nucleon with momentum $p^{\prime}$ is taken on shell
($(p^{\prime})^2=m^2$, $m$ being the nucleon mass) and       
the initial  one with momentum $p$ is off-shell. The operator
$\Lambda_{+}(p)$ is defined according to
\beq
\Lambda_{\pm}(p) \equiv \frac{{\pm}\vslash{p}+m}{2m} \;.
\eqlab{proj}
\eeq
The vertex functions $G_{ps}(p^2)$ and $G_{pv}(p^2)$
denote the form factors corresponding to the usual pseudo-scalar and
pseudo-vector couplings.                                         
The renormalized dressed nucleon propagator can be written as,
\beq
S(p)=\left[\,Z_2\, (\vslash{p}-m_0)-A(p^2) \vslash{p} -
B(p^2) m+i0 \,\right]^{-1},
\eqlab{n-prop}
\eeq
where the nucleon self-energy functions $A(p^2)$ and $B(p^2)$ have been
introduced together with the renormalization constant $Z_2^N$ and the bare
nucleon mass $m_0$.

Imposing the condition of analyticity of the amplitude at the level of
one-particle reducible 4-point diagrams, the dressing is formulated in
terms of a set of coupled equations, schematically shown in
\figref{piNN}. These express the dressing of the vertex and the
propagator by multiple meson loops, chosen such as to be consistent with
the subsequent application in the K-matrix calculation, discussed later.
The K-matrix procedure generates imaginary parts of the T-matrix,
necessary for obeying unitarity relations. However, these imaginary
parts are not related to the real parts as one would expect for an
analytic function (Cauchy's theorem will be violated). The vertex and
self-energy corrections generated in the present approach are such that
analyticity relations are restored while still maintaining the unitarity
constraints. The reader is referred to~\cite{Kon99,Kon00a} for a full
account of the procedure used.
}

The equations presented in \figref{piNN} are solved by iteration where
every iteration step proceeds as follows. The imaginary -- or pole --
contributions of the loop integrals for both the propagators and the
vertices are obtained by applying cutting rules \cite{Man59}. Since the
outgoing nucleon and the pion are on-shell, the only kinematically
allowed cuts are those shown in \figref{piNN}. The real part
of the vertex (i.e. the real parts of the form factors) 
and self-energy functions
are calculated at every iteration step by applying dispersion relations
\cite{Bin60} to the imaginary parts just calculated, where only the
physical one-pion--one-nucleon cut on the real axis in the complex
$p^2$-plane is considered. These real functions are used to calculate
the pole contribution for the next iteration step. This procedure is
repeated to obtain a converged solution. We consider irreducible
vertices, which means that the external propagators are not included
in the dressing of the vertices.

One of the advantages of the use of cutting rules in the solution procedure
is that throughout we deal with vertices with only one virtual nucleon
(half-off-shell vertices). In other words, the knowledge of {\it
full}-off-shell form factors will not be needed for the calculation of
the pole contributions to the loop integrals,
which greatly
simplifies our approach. Also for the construction
of the K-matrix only half-off-shell vertices are required.

In the dressing procedure we maintain the full Lorentz structure of the
nucleon propagator and $\pi N N$ vertices.
The half-off-shell $\pi N N$ vertex, as 
given in Appendix B, is written in terms of the form factors 
$G_{ps}(p^2)$ and $G_{pv}(p^2)$ corresponding to the usual pseudo-scalar and
pseudo-vector couplings.

Bare $\pi NN$ form factors
\beq
G_{pv}^0(p^2)=f\,(1-\chi)\,
\exp{\left[-\ln{2}\frac{(p^2-m^2)^2}{\Lambda_N^4}\right]}
 \mbox{\ \ \  and \ } G_{ps}^0(p^2)=\frac{\chi}{(1-\chi)} \,G_{pv}^0(p^2)
\eqlab{pinn_bare}
\eeq
have been introduced in the dressing procedure to regularize the
dispersion integrals. Here $\Lambda_N^2$ is the half-width of the form factor, and
$f$ is a bare coupling constant fixed from the condition that the
dressed vertex reproduces the physical pion-nucleon coupling on-shell.
For simplicity we have taken a vanishing pseudo-scalar
admixture  ($\chi=0$) in the bare vertex. The bare form factor reflects
physics at energy scales beyond those of the included mesons and which 
has been left out of the dressing procedure. One thus expects a
large width for this factor, as is indeed the case. The use of a bare
form factor also implies a violation of analyticity due to additional
singularities of this form factor. However, since the
width of the form factor is large, the associated violation of
analyticity will be small in the energy regime of present interest. It
should be noted that our results are largely insensitive to the details
of the structure of the bare form factor; only its width matters.

In dressing the $\pi N N$ vertex, the $\Delta$, $\rho$ and $\sigma$
degrees of freedom are taken into account besides the pion (the vertices
and parameters are given in Appendix B and in \tabref{mesons}). The
coupling parameters were adjusted so that the iteration procedure
converges and also to get a reasonable reproduction of pion-nucleon
phase shifts in the full model calculation, discussed in Section III B.
We have insisted on consistency between model parameters in the
calculation of the vertices and in the full model calculation. For
reasons of simplicity, however, we have not included the full nucleon
resonance spectrum in the dressing of the vertices; nor have we 
dressed all resonance propagators and vertices on the same footing with the 
nucleon propagator and $\pi N N$ vertex. This can be considered as an additional
approximation in the present approach.

The dressed nucleon propagator is renormalized to have a pole with a unit
residue at the physical mass and is given in 
Appendix B, where the nucleon self-energy functions $A(p^2)$ and $B(p^2)$ 
have been
introduced together with the renormalization constant $Z_2^N$ and the 
nucleon mass shift $\delta m=m-m_0$.

\subsubsection{The $\gamma$NN vertex}
\seclab{gnn}

The procedure of obtaining the $\gamma N N$ vertex is in principle the same 
as for the $\pi N N$ vertex. One should consider the cut loop diagrams generated
in the K-matrix approach and use the equivalent cuts for evaluating the 
integrand of the dispersion integral. This is shown schematically in 
\figref{gamNN}, a more complete discussion is presented in~\cite{Kon00b}. 
This equation is solved using the same method as used for constructing
the $\pi N N$ vertex.
The equation in \figref{gamNN} is simpler than that in \figref{piNN} since, 
due to the weaker electromagnetic coupling, 
photon loops are not considered. The dressed $\pi N N$ vertex can thus be 
taken from the calculations presented in the previous section.

The most general $\gamma NN$ vertex is kept in the
dressing procedure. It is given in Appendix B 
for a real photon with momentum
$q=p'-p$ and an on-shell outgoing nucleon,
$p^{\prime 2} = m^2$.
It contains four form factors $\hat{F}^{+,-}_{1,2}(p^2)$ each of which has
the isospin structure $\hat{F}=F^s + \tau_3 F^v$.

\Omit{
The most general $\gamma NN$ vertex for a real photon with momentum
$q=p'-p$, in which the outgoing nucleon is on the mass-shell,
$p^{\prime 2} = m^2$, can be written as 
\beq
\Gamma_{\mu}(p) = -i\,e \sum_{l={\pm}} \left[ \gamma_{\mu} \hat{F}_1^{l}(p^2) +
i \frac{\sigma_{\mu \nu}q^{\nu}}{2m} \hat{F}_2^{l}(p^2)
\right]\,\Lambda_l(p)\;,
\eqlab{finon}
\eeq
where $e$ is the elementary electric charge
of the nucleon and the operators $\Lambda_l(p)$ have been defined in
\eqref{proj}. The isospin structure of the form factors is taken as
$\hat{F}=F^s + \tau_3 F^v$.

Once the dressed $\pi$NN vertex has been calculated from the equations
in \figref{piNN}, the $\gamma NN$ vertex follows from the much simpler
recursive equation \cite{Kon00b} shown schematically in \figref{gamNN}.
This equation is solved using the same method as used for constructing
the $\pi$NN vertex.

The bare $\gamma N N$ vertex is taken as 
\beq
\Gamma_{\mu}^0(p) =
-i\,e\,\left( \gamma_{\mu} \hat{e}_N + i \hat{\kappa}_B
\frac{{\sigma}_{\mu \nu} q^{\nu}}{2 m} \right),
\eqlab{ver_bare}
\eeq
where $\hat{e}_N = (1+{\tau}_3)/2$ and $\hat{\kappa}_B = \kappa_B^s
+ \tau_3 \kappa_B^v$ is the bare anomalous magnetic moment of
nucleon, adjusted to provide the normalization
\beq
F_2^{+,s}(m^2)=-0.06 \mbox{\ \ \ and\ \ }
F_2^{+,v}(m^2)=1.85 \;.
\eqlab{normalF2}
\eeq
of the dressed vertex. We have not introduced form factors in
\eqref{ver_bare} since the dispersion integrals are finite due to the
sufficiently fast falloff of the dressed $\pi N N$ vertex.
}

The bare $\gamma N N$ vertex is taken with 
$\hat{F}^{+}_{1}(p^2)=\hat{F}^{-}_{1}(p^2)=\hat{e}_N = (1+{\tau}_3)/2$ 
and 
$\hat{F}^{+}_{2}(p^2)=\hat{F}^{-}_{2}(p^2)=\hat{\kappa}_B = \kappa_B^s
+ \tau_3 \kappa_B^v$, 
the bare anomalous magnetic moment of
the nucleon, adjusted to provide the normalization
\beq
F_2^{+,s}(m^2)=-0.06 \mbox{\ \ \ and\ \ }
F_2^{+,v}(m^2)=1.85 \;.
\eqlab{normalF2}
\eeq
of the dressed vertex. We have not introduced bare $\gamma N N$ form factors 
since the dispersion integrals are finite due to the
sufficiently fast falloff of the dressed $\pi N N$ vertex.

The contact $\gamma \pi N N$ and $\gamma \gamma N N$ vertices, 
necessary for gauge invariance of the model, are constructed by minimal
substitution in the dressed $\pi N N$ vertex and nucleon propagator, as
was explained in \cite{Kon00b}. Using minimal substitution, two $\gamma
\pi N N$ vertices were derived in \cite{Kon00b}, which differ by a
purely gauge invariant (transverse with respect to the photon
four-momentum) term. In the present calculation, the contact term was
chosen as a weighted sum of the two $\gamma \pi N N$ vertices, 
\beq
\left( \Gamma_{\gamma \pi N N}^{ps}\right)_\alpha^\mu+
\left[0.15\, \left( \,\Gamma_{\gamma \pi N N}^{pv \,1}\right)_\alpha^\mu+
0.85\, \left( \Gamma_{\gamma \pi N N}^{pv \,2}\right)_\alpha^\mu\, 
\right] \,,
\eqlab{ct_res:ch5}
\eeq
using the notation introduced in Eqs.~(C.3,C.5) and (C.6) of Ref.~\cite{Kon00b}.
The reason for this choice will be discussed in \secref{observ},
when presenting results for pion photoproduction.
Since, due to the inclusion of the $\gamma \pi N N$ contact term,
the photon vertex  obeys the Ward-Takahashi identity,
the $F_1^{\pm}$ form factors are uniquely related to the nucleon self-energy
(see Ref.\cite{Kon00b}).

As explained in \secref{analyt}, the present procedure restores 
analyticity at the level of one-particle reducible diagrams in the T-matrix.
In general, 
violation due to two- and more-particle reducible diagrams can be 
regarded as higher order corrections.
An important exception to this 
general rule is formed by one-particle irreducible diagrams 
generated by the K-matrix 
formalism for Compton scattering.
These include, for example, diagrams where both photons couple to the same 
intermediate pion in a loop (so-called ``handbag" diagrams).
In the K-matrix formalism, 
the imaginary (pole) contribution of this type 
of diagrams is essentially a square of part 
of the pion-photoproduction amplitude. This term is exceptional since at 
the pion threshold the S-wave contribution is large, 
related to the non-zero value of 
the $E_{0+}^{1/2}$ multipole in pion-photoproduction. 
Not including the real part of such a large contribution would
entail a significant violation of analyticity.
To correct for this to a large extent, 
the $\gamma \gamma N N$ vertex also contains the (purely transverse) 
``cusp" contact term whose construction is
described in Section 4 of Ref.~\cite{Kon00b}. It is derived from a 
simplified treatment of the large ``handbag" 
loop correction, using the fact that the latter is dominated by the 
$E_{0+}^{1/2}$ photoproduction channel and that therefore the loop can 
be saturated by $J^\pi=1/2^-$ intermediate states. 
Since these intermediate states 
correspond to negative-energy states, the integrand of 
the dispersion integral for the ``cusp" term is 
proportional to the negative-energy {\it magnetic} (to ensure the
transversality of the term) form factor as calculated 
in the dressing procedure. 
The explicit expression for the term is given in
Eqs.~(32) and (33) of Ref.~\cite{Kon00b}, where the functions 
$\mbox{Im} \widetilde{F}_{2}^{-}$ and
$\widetilde{\Sigma}_I$ are now obtained from the full dressing procedure.

Since, due to chiral symmetry, the S-wave pion scattering amplitude 
vanishes at threshold, $\pi\pi N N$ or $\pi \gamma N N$ contact 
terms analogous to the ``cusp" $\gamma \gamma N N$ term 
will be negligible and have therefore not been considered.

\Omit{
The $\gamma \gamma N N$ vertex also contains the ``cusp" contact
term\cite{Kon00b}, where the functions $\mbox{Im} \tilde{F}_{2}^{-}$ and
$\tilde{\Sigma}_I$ are now obtained from the full dressing procedure.
}

\subsection{The effective Lagrangian}
\seclab{ver_vec:ch5}

Since our approach is based on the K-matrix formalism, it is convenient
to describe  the effective Lagrangian of our model in terms of
the diagrams included in the kernel $K$.

The $S$ matrix is expressed in terms of the scattering amplitude
$\mathcal{T}$ (the T matrix) by $ S=1+2i{\mathcal{T}}$. The T-matrix is
written in the form $ {\mathcal{T}}\,=\,K\,+\,K\,i\delta\,{\mathcal{T}}
$, which can be solved, yielding the central equation of the K-matrix
approach,
\beq
{\mathcal T}= (1- K i\delta)^{-1}\,K \; ,
\eqlab{T-K}
\eeq
where $\delta$ indicates that the intermediate particles have to be
taken on the mass shell.
It is straightforward to check that $S$ is unitary provided $K$ is
hermitian. The simplicity of \eqref{T-K} is due to the fact that
it contains the cut two-particle propagator $i\delta$, 
thus involving integrals only over on-shell intermediate
particles. As a result
\eqref{T-K} reduces to a set of
algebraic equations when one is working in a partial wave basis.
When both the $\pi - N$ and $\gamma - N$ channels are open, the
coupled-channel K-matrix becomes a $2 \times 2$ matrix
in the channel space, i.e.\ 
\beq
K=\left[
\begin{array}{cc}
 K_{\gamma \gamma} & K_{\gamma \pi} \\
 K_{\pi \gamma} & K_{\pi \pi}
\end{array}
\right].
\eqlab{k-matr}
\eeq
The pion-nucleon scattering entry $K_{\pi \pi}$ of the K matrix
is given in \figref{K-pipi}, and the
pion photoproduction and Compton scattering entries, $K_{\pi \gamma}$ and
$K_{\gamma \gamma}$,
are shown in Figs.~(\ref{fig:K-gampi}) and (\ref{fig:K-gg}),
respectively, including all baryon resonances specified 
in \tabref{reson}.

The analytic form of the K matrix can be written by applying Feynman rules,
with the 3-point vertices and propagators given in Appendix B.

Several coupling constants are fixed from the corresponding decay
widths, according to the Particle Data Group values \cite{Gro00}:
$g_{\rho^0 \pi \gamma}=0.131$, $g_{\rho^\pm \pi \gamma}=0.103$ and
$g_{\omega \pi \gamma}=0.313$. The coupling constant $g_{\pi \gamma
\gamma}$ is fixed so that the width of the pion decay into two photons
is 7.37(1.5) eV. The remaining parameters, given in
\tabref{mesons} for the nucleon-meson couplings and in \tabref{reson}
for the coupling to the resonances, were determined from a fit to
pion-nucleon phase shifts, multipole amplitudes in pion photoproduction and
Compton scattering cross sections. The range of the parameters
pertaining to the degrees of freedom included in the dressing equations
is constrained in addition by the requirement of convergence of the
iteration procedure\cite{Kon00a}.

\Omit{
In the present calculation only tree-level diagrams are included in the kernel. 
Partly for 
reasons of simplicity, we have not included more complicated diagrams 
(in addition to those generated by the dressing procedure). 
Effects of these diagrams can be captured 
Another reason 
for not doing so is that the effects of most more complicated diagrams 
can be captured by a more complicated dependence on momenta of the structures of 
the different vertices which are already incorporated, Since in the 
present paper we do not aim for a perfect fit to the data we have not 
investigated this in detail. In addition we have not performed a 
least-square fit of the present parameters. 
}

\subsection{Violation of analyticity as an expansion parameter} 
\seclab{analyt}

As we have argued, the amplitude in this model obeys crossing symmetry
at the expense of violating analyticity. We have incorporated
analyticity constraints at the level of 1-particle reducible diagrams
contributing to the amplitude. Analyticity is violated at the level of
2- and higher-particle reducible diagrams (the higher-particle reducible
diagrams enter when multi-pion production channels are included
explicitly). Our approach can thus be regarded as an expansion in a
certain parameter $a$, the degree of analyticity of the amplitude, where
we interpret the real (principal-value) parts of the dressed
one-particle reducible diagrams as terms of ${\mathcal{O}}(a)$, the real
parts of the dressed 1-particle irreducible but 2-particle reducible
diagrams as $\sim {\mathcal{O}}(a^2)$ and so on. Without specifying it,
the expansion parameter $a$ is related to the energy scale at which the
amplitude is affected significantly by imposing the analyticity
constraints. By phase-space arguments, the higher-particle reducible
diagrams will have finite and comparatively smooth imaginary parts only
at higher energies. The associated real parts will thus exhibit an
energy dependence which is weaker than those due to the 1- and
2-particle reducible diagrams. In the approach presented here, these
weak energy dependences are in general absorbed in one of the fitting
constants of the model, such as the strength of the negative energy
couplings of the resonances.

We will show in \secref{vert} that analyticity at the one-body level
(order ${\mathcal{O}}(a)$) strongly affects 3-point vertex functions at
relatively moderate momenta. The approach can be extended to include
also analyticity at the level of 4-point 2-particle reducible
functions. To do so, one will have to add 4-point contact term
contributions to the K-matrix. Note that the corresponding imaginary
parts are associated with the one-pion--one-nucleon discontinuity. Other
diagrams of the orders higher than ${\mathcal{O}}(a^2)$ will be
associated with one-nucleon--multi-pion thresholds, which implies that
they become important above energies of order $m_N+2\,m_\pi$. In other
words, in the approximation of two-body unitarity, only the terms of
${\mathcal{O}}(a)$ and ${\mathcal{O}}(a^2)$ should be retained. Thus, to
systematically improve the property of analyticity of the amplitude, one
has to extend the dressing technique to the 1-particle irreducible (but
2-particle reducible) 4-point diagrams. This will ensure that all the
terms up to and including $\sim {\mathcal{O}}(a^2)$ are taken into
account. To go beyond second order, one would have to accommodate
two-pion production in the model.

In the case of Compton scattering, the ``handbag'' diagram gives a large
contribution with pronounced features at the pion production threshold.
Therefore, although this diagram enters at ${\mathcal{O}}(a^2)$, we
had to include it effectively through the ``cusp'' contact
term\cite{Kon00b}.

An additional breaking of analyticity is caused by the introduction
of the bare form factors. Since these are wide, the associated poles in
the complex $p^2$-plane will be far removed from our region of interest
and one may thus regard the associated violation of analyticity being of
higher order in $a$. Since a form factor is associated with physics not
included explicitly in the model, which in the present case is multi-pion
production, one could argue that the associated breaking is of the order
${\mathcal{O}}(a^3)$.

\section{Results }  \seclab{pi-N}

Results obtained from the dressed K-matrix model are presented in this
section. We first discuss effects of the dressing for the vertex
functions and nucleon self-energy, showing the large effect of
multi-loop dressing. Since these off-shell form factors are not
observable quantities, results for pion-nucleon scattering, pion
photoproduction and Compton scattering are presented in the following
sections.

\subsection{Vertex functions}   \seclab{vert}

As explained above, the dressing of the vertices is expressed in terms
of form factors and self-energy functions. They depend of the choice of
the bare form factors, but our calculations show that the detailed
structure of this bare form factor is rather unimportant, only its
half-width $\Lambda_N^2$ is crucial. There exists a maximum width beyond
which the dressing procedure fails to converge. This maximum depends on
the meson-nucleon couplings and on the pseudo-scalar -- pseudo-vector
mixing ratio ($\chi$) used in the bare form factor. The width we have
taken in the calculations discussed in the following, see
\tabref{mesons}, has been taken relatively close to this maximum. The
form factors for the pion-nucleon vertex are shown in \figref{f-pi}.

Due to the dressing, the pion-nucleon vertex is considerably softer than
the original bare vertex. The difference is about a factor one and a half 
in the width. The dressing affects the pseudo-scalar and pseudo-vector parts of
the vertex function differently, resulting in a mixing ratio which is
strongly momentum dependent. At the pion-production 
threshold the ratio is still small, i.e.\ the pseudo-vector structure dominates,
which is consistent with a minor explicit breaking of chiral symmetry.

The nucleon self-energy functions are shown in \figref{n-self}. The wave
function and mass renormalization constants are
$Z_2^N=0.8$ and $\delta m=-77$ MeV, respectively. The bare pion coupling
constant is $f=10.82$ in \eqref{pinn_bare}. In principle, a field
redefinition can be made such that the self-energy vanishes, resulting
in a transformation of the $\pi N N$ vertex (see \cite{Kon00a,Kon01} for
details). Due to the equivalence theorem\cite{Chi61}, observables
calculated in both representations are the same. In the present
calculation we have not used a field redefinition. 

Since the model is gauge invariant -- and thus the $\gamma N N$ vertex
obeys the Ward-Takahashi identity -- there is a one-to-one
correspondence between the electric form factors $F_1^{\pm}$ and the
nucleon self-energy, given by Eqs.~(28) and (29) of Ref.~\cite{Kon00b}. 
In particular, the neutron-photon electric form
factors are zero. The proton-photon form factors $F_1^{\pm}(p^2)$ are
shown in \figref{f1-p} as functions of the momentum squared of the
proton. They do not depend on the choice of the contact $\gamma \pi N N$
vertex.

The magnetic form factors $F_2^{\pm}$ are shown in \figref{f2} for the
proton and the neutron. The dominant contribution to the form factors
$F_2^{\pm}$ is due to the first diagram under the integral in
\figref{gamNN}, which generates the bulk of the form factors already in
the first iteration. This, however, does not mean that the other terms
in the equation are of minor importance. In particular, they are crucial
for satisfying the Ward-Takahashi identity for the $\gamma N N$ vertex.
Because the derivative of the imaginary part of $F_2^{-}(p^2)$ tends to
infinity as $p^2$ approaches the pion threshold (since intermediate
s-wave $\pi - N$ states give a large contribution to the loop integrals in
\figref{gamNN}), its real part has a sharp cusp. The dressed vertex is
renormalized by adjusting $\hat{\kappa}_B$ ($\kappa^s_B=0.017$ and
$\kappa^v_B=1.78$) in the bare $\gamma N N$ vertex to fulfill the
normalization conditions \eqref{normalF2}.

\subsection{Observables in pion-nucleon scattering and pion photoproduction}
\seclab{observ}

In the following we shall be discussing effects of the dressing of
vertices and propagators on observables by comparing two calculations,
referred to as calculations B (Bare) and D (Dressed).

\begin{itemize}
\item
{\it Calculation B}.
The ``bare" K matrix, $K_B$, consists of free propagators and bare
vertices for all particles. No form factors are included, except for the
bare form factors in the $\pi N N$ and $\pi N \Delta$ vertices. A
$\gamma \gamma N N$ term is not included since the bare $\gamma N N$
vertex does not have form factors. Thus, the corresponding T matrix
contains only the pole parts of the loop diagrams.
\item
{\it Calculation D}.
The ``dressed" K matrix, $K_D$, is composed of the dressed propagators
for the nucleon, $\Delta$, $\rho$ and $\sigma$ (the propagators of the
pion and $\omega$ are taken free) and the dressed $\pi N N$ and $\gamma
N N$ vertices. To provide gauge invariance of the Compton scattering
amplitude, a $\gamma \gamma N N$ vertex is added, which includes the
additional ``cusp" term. Since this calculation includes all features of
the model, it is also referred to as the full calculation. Now both pole
and principal-value parts are taken into account of a wide class of loop
diagrams which contribute to the T matrix. \end{itemize}

\noindent Both calculations include a contact $\gamma \pi N N$ vertex,
calculated by minimal substitution in the $\pi N N$ vertex (in the bare
vertex for calculation B and in the dressed one for calculation D). The
$\pi N N$ and $\gamma N N$ vertices are normalized at the threshold to
reproduce the physical pion-nucleon coupling constant and the nucleon
anomalous magnetic moment, respectively. In addition, nucleon resonances
are included in both calculations. Since we focus on effects of the
nucleon dressing, we do not readjust the resonance parameters in
calculation B.

The pion-coupling parameters have been optimized to reproduce the
pion-nucleon scattering phase shifts and inelasticities in the full
calculation. However, since in the present paper we do not aim for a
perfect fit to the data,
we have not used a least-square minimization routine.
In part, this is because certain phase shifts, 
notably the $P_{11}$ at higher energies,
show discrepancies which seem to be outside the capability of the model.
On the whole, the phase shifts and inelasticities are reproduced well,
see \figref{pi-pi}. In the propagators for the nucleon resonances an
additional width $\Gamma_0$ has been introduced, given in
\tabref{reson}, to account for decay channels which are not included
explicitly in the model (see Ref.\ \cite{Kor98} for details). The large
effect of the dressing which was seen in the vertex functions persists in
the calculation of observables. At pion energies exceeding 500 MeV the
phase shift in the $P_{33}$ channel is somewhat above the data. This
appears to be due to the structure of the $\pi N \Delta$ vertex. Namely,
due to the use of a gauge-invariant $\pi N \Delta$ vertex \cite{Pas98},
an additional factor $p^2$ is introduced in the s-channel diagrams,
which is apparently insufficiently compensated by the form factor. We
have not made an extensive effort to correct this in the present
calculation.

The calculated multipoles for pion photoproduction on the proton are shown in
\figref{pi-gam}, where the results of both calculations D and B are given.
The usual nomenclature \cite{Gar93} for the multipoles is used.
Comparing the results of the two calculations in \figref{pi-gam},
it is seen that effects of the dressing are most prominent in the
magnetic dipole multipole $M_{1-}^{1/2}$ reflecting the effect of the
$F_2^+$ form factors in the s-channel.

As pointed out earlier, gauge invariance alone does not provide
sufficient restrictions on the $\gamma \pi N N$ contact term: its
transverse part cannot be determined unambiguously. We
found\cite{Kon00b} that the choice of the transverse part has an
influence on the multipole $E_{0+}^{1/2}$, in particular its falloff
with energy, which allowed us to fix this term as given in
\eqref{ct_res:ch5} to fit the data. Since
the $E_{0+}^{1/2}$ multipole corresponds to angular momentum and parity
$J^\pi=1/2^-$ of the nucleon-photon system, it contributes to the
imaginary part of $F_2^-$, which explains the strong effect of the
$\gamma \pi N N$ contact term on $F_2^-$.

\section{Compton scattering and nucleon polarizabilities} \seclab{ComPol}

Our special interest concerns observables in Compton scattering since
for this case constraints imposed by crossing symmetry and analyticity
will be most important.

\subsection{Compton scattering}

The only two free parameters which enter in the calculation of Compton
scattering appear in the $\gamma\gamma\sigma$ vertex and have been
adjusted to reproduce the backward Compton cross section at moderate
energies. Calculated angular distributions are compared with
data in \figref{ggs}. Notably,
the near vanishing $0^o$ cross section at the pion production energy is
correctly reproduced in the full calculation, a feature which is 
impossible to obtain in the
calculation without dressing. At somewhat higher energies, however, the
forward cross section is under-predicted in the full calculation.
The dependence of the cross section on the photon energy is displayed in 
\figref{cs_en} for different scattering angles. In general
the full calculation gives an improvement in the description of the data.

We have not made a detailed investigation of the 
source of the observed discrepancies with the cross section data.
Below the energy of the $\Delta$ resonance the contribution of $\sigma$-meson 
exchange is small and the most obvious reason for the 
problems would be the structure of the $\Delta N \gamma$ vertex.
At smaller 
angles there is destructive interference between the $\Delta$ and the 
nucleon contribution. A relatively minor change in the momentum dependence of
the $\Delta N \gamma$ vertex will 
thus be magnified in its effect on the cross section at forward angles 
while affecting the larger angles to a lesser extend.
For example, the poor agreement at backward angles could be mitigated by
lowering the absolute value of the mixing parameter $a_\gamma$ in
the $\Delta N \gamma$ vertex (see Appendix B and Table II), which would 
however result in a too high cross section at smaller angles.
Above the energy of the $\Delta$ resonance, the contribution of
the $\sigma$ meson becomes progressively more important. As a t-channel 
exchange, it mainly affects the backward angles. Thus, at higher energies and 
backward angles the cross 
section will be sensitive to the structure of the vertices 
in the $\sigma$ exchange diagram. 
The detailed 
study of such modifications to the $\Delta$ and $\sigma$ vertices 
falls, however, outside the scope of the present work.

\Omit{
The Compton cross section is shown in the upper panel of
\figref{gam-qnp} as a function of the photon laboratory energy, at a
scattering angle of $90^o$.
}
The photon (beam) asymmetry
at $90^o$ and the proton (target) polarization at $100^o$ are shown in
\figref{gam-qnp} as functions of the photon laboratory energy. As can be
seen, effects of the dressing become very
conspicuous above the $\Delta$ resonance region. The observables exhibit
a cusp structure at the pion threshold, which is especially pronounced
for the photon asymmetry. This cusp is a consequence of the unitarity
and analyticity properties of the coupled-channel scattering matrix,
affecting primarily the $f_{EE}^{1-}$ partial wave amplitude 
\cite{Ber93,Kon00b}.

The effect of the dressing on the $f_{EE}^{1-}$ amplitude can be seen in
\figref{fee}, where also the results of dispersion analyses are quoted
for comparison. Note that the imaginary parts of $f_{EE}^{1-}$ from
calculations B and D are rather similar in the vicinity of threshold.
Both calculations B and D are unitary, and the full calculation D
includes the dressing in addition. Since gauge invariance, crossing and
CPT symmetries are fulfilled in the model, all these calculations tend
to the Thompson limit at vanishing photon energy.

\subsection{Nucleon polarizabilities}

The polarizabilities characterize response of the nucleon to an
externally applied electromagnetic field \cite{Hem98,Hol00}. They can be
defined as coefficients in a low-energy expansion of the cross section
or partial amplitudes of Compton scattering. We adhere to the standard
notation for the partial amplitudes \cite{Pfe74,Ber93}. We will attach
the superscript $NB$ (non-Born) to the difference between the full
amplitude obtained in the full calculation D and the amplitude in the
Born approximation. In the Born approximation, the T matrix equals the
sum of the first two graphs in \figref{K-gg} with the bare (but properly
normalized to the physical anomalous moment) $\gamma N N$ vertices and
the free nucleon propagator. Such a calculation is not unitary,
resulting in a purely real amplitude. According to the low-energy
theorem~\cite{Low54}, the zeroth and first orders in an expansion of the
amplitude in the small photon energy $\omega$ are model-independent and
are reproduced by the Born contribution alone. The polarizabilities
enter starting at second order and are model-dependent. We are in
particular interested in the role of the dressing procedure in this
connection.

To calculate the polarizabilities, we use formulae given in Ref.~\cite{Bab98}.
The electric and magnetic (scalar) polarizabilities are determined
using the equations
\beq
\alpha_{E} \simeq \frac{(f_{EE}^{1-}+2 f_{EE}^{1+})^{NB}}{\omega^2},\;\;\;
\beta_{M} \simeq \frac{(f_{MM}^{1-}+2 f_{MM}^{1+})^{NB}}{\omega^2}.
\eqlab{al_bet}
\eeq
The spin (vector) polarizabilities are
related to third order coefficients in the low-energy expansion,
\beq
\gamma_{E1} \simeq \frac{(f_{EE}^{1+}-f_{EE}^{1-})^{NB}}{\omega^3},\;\;\;
\gamma_{M1} \simeq \frac{(f_{MM}^{1+}-f_{MM}^{1-})^{NB}}{\omega^3},
\eqlab{gam_em1}
\eeq
\beq
\gamma_{E2} \simeq \frac{6 (f_{ME}^{1+})^{NB}}{\omega^3},\;\;\;
\gamma_{M2}\simeq \frac{6 (f_{EM}^{1+})^{NB}}{\omega^3}.
\eqlab{gam_em2}
\eeq
We calculate also the forward- and backward-angle spin polarizabilities,
given by
\beq
\gamma_0=-\gamma_{E1}-\gamma_{M1}-\gamma_{E2}-\gamma_{M2}
\;\;\;\; \mbox{and} \;\;\;\;
\gamma_\pi=-\gamma_{E1}+\gamma_{M1}+\gamma_{E2}-\gamma_{M2},
\eqlab{forback}
\eeq
respectively. We obtained similar values for the polarizabilities
extracted at the energies in the range between $\omega=20$ MeV and
$\omega=100$ MeV, whereas at lower energies the numerical extraction was
unreliable due to the closeness of the amplitude to the nucleon pole.
For this reason, we applied a linear extrapolation to $\omega=0$ MeV
through the values of polarizabilities calculated at $\omega=80$ MeV and
$\omega=40$ MeV (the points used for the extrapolation are immaterial
within the indicated range of moderately low energies).

Since all the parameters are now fixed by the dressing procedure and by the
comparison with experiment for the pion-nucleon scattering, pion
photoproduction and Compton scattering, the calculated polarizabilities
reflect the dynamical contents of the model.
Our results for the electric, magnetic and spin polarizabilities of the
nucleon are given in Tables~\ref{tab:polp} and ~\ref{tab:poln}, where
also results of other calculations are summarized together with the
values extracted from recent experiments. (It should be noted that there
has been some discussion concerning the definition of the
polarizabilities used in different
chiral perturbation theory calculations, see \cite{Bir00}.) 

It is known~\cite{Bab98} that the t-channel $\pi^0$-exchange diagram
gives a large contribution to the spin polarizabilities $\gamma$, while
not affecting the scalar polarizabilities $\alpha$ and $\beta$. For this
reason, this contribution is often subtracted from the $\gamma$s, as is
also done in Tables~\ref{tab:polp} and~\ref{tab:poln}. We find that the
$\pi^0$-exchange diagram gives a contribution of $+10.62$ to
$\gamma_{E1}^p$, $\gamma_{M2}^p$, $\gamma_{M1}^n$, $\gamma_{E2}^n$ and
$-10.62$ to $\gamma_{E1}^n$, $\gamma_{M2}^n$, $\gamma_{M1}^p$,
$\gamma_{E2}^p$ (earlier works quote similar numbers: $\pm 11.3$
\cite{Ber95} $\pm 11.2$ \cite{Lvo97}, $\pm 10.9$ \cite{Hem98}, $\pm
10.7$ \cite{Vij00}). The effect of the dressing on the polarizabilities
can be seen by comparing the values given in columns D and B. In
particular, the dressing tends to decrease $\alpha$ while increasing
$\beta$. Among the spin polarizabilities, $\gamma_{E1}$ is affected much
more than the other $\gamma$s.

Various contributions to the full calculation D of the polarizabilities
are analyzed in Tables~\ref{tab:polp-contr} and \ref{tab:poln-contr} for
the proton and neutron, respectively. The different rows contain the
results obtained from the calculation in which certain contributions
have been omitted. The $\Delta$-resonance, with its strong magnetic
coupling, primarily affect the magnetic polarizabilities such as
$\beta$, $\gamma_{M1}$ and $\gamma_{M2}$.

The $\sigma$ meson does not affect the sum of the scalar
polarizabilities, $\alpha+\beta$. To understand this feature, we recall
that the second order term in the low-energy expansion of the
differential cross section in the laboratory frame of reference can be
expressed in terms of the polarizabilities as \cite{Gui78}
\beq
-\frac{m}{2\alpha_f}\Big[(\alpha+\beta)(1+cos \theta)^2+
(\alpha-\beta)(1-cos \theta)^2\Big]\,\omega^2,
\eqlab{c2_pol}
\eeq
$\alpha_f$ being the fine structure constant, $\alpha_f=1/137$. The sum
$\alpha+\beta$ thus remains unaffected since $\sigma$-exchange enters as
a t-channel contribution and does not contribute at forward angles. Also
the spin-polarizabilities are not affected by a scalar exchange. Both
the $\Delta$ and the $\sigma$ give large, but cancelling, contributions
to $\beta$.

The effect of the additional ``cusp'' $\gamma \gamma N N$ contact
term\cite{Kon00b}, mentioned at the end of \secref{gnn}, can be
seen by comparing the ``no cusp'' with the full calculation. In
particular, it is seen that this term strongly influences the electric
polarizabilities rather than the magnetic ones. The reason for this is
that the ``cusp" contact term affects primarily the electric partial
amplitude $f_{EE}^{1-}$ (corresponding to the the total angular momentum
and parity of the intermediate state $J^\pi=1/2^-$) rather than the
magnetic amplitude $f_{MM}^{1-}$ ($J^\pi=1/2^+$). Hence, by
Eqs.~(\ref{eq:al_bet},\ref{eq:gam_em1}), the electric polarizabilities
($\alpha$ and $\gamma_{E1}$) receive a sizable contribution from this
term.

None of the polarizabilities is much influenced by contributions from
the $D_{13}$ or any of the other resonances.

\section{Summary and Conclusions}

The results are presented of a comprehensive calculation of pion and
photon scattering off the nucleon in the ``Dressed K-matrix Model". In
particular, we focused our attention on the calculation of Compton
scattering in the energy regime ranging from the lowest energies, where
observables are presented in terms of nucleon polarizabilities, up to
energies in the second resonance region. We show that this model
distinguishes itself from other microscopic approaches since it is able to
give a quantitative description of the observables in the full energy
range, due to the fact that the model preserves the symmetries which are
important in the different energy regimes.

\acknowledgments

We would like to thank Alex Korchin for discussions.
We also thank Harold Fearing for reading the manuscript and making
several useful suggestions.
This work is part of the research program of the ``Stichting voor
Fundamenteel Onderzoek der Materie'' (FOM) with financial support
from the ``Nederlandse Organisatie voor Wetenschappelijk
Onderzoek'' (NWO).

\appendix

\section{Crossing symmetry}

The proof that our T-matrix obeys crossing symmetry in the meson lines
is based on simple kinematics. Diagrams of the type depicted on the left
in \figref{diag-d} contribute to the T-matrix. Crossing symmetry of the
T-matrix implies that also the crossed version of this diagram, depicted
on the right in \figref{diag-d}, is taken into account. Part of this
diagram corresponds however to the incoming nucleon  
``decaying" to a
state consisting of two on-shell pions and an on-shell nucleon. Since
this is not allowed kinematically, the contribution from this crossed
diagram vanishes.

This argument can be used to show that for any term contributing to the
T-matrix which involves a second or higher power of the kernel, the
corresponding crossed diagram vanishes. The resulting T-matrix is thus
crossing symmetric provided that the kernel itself is crossing
symmetric. In the present approach special care is taken that the latter
is indeed the case.

\section{Model Lagrangian}

The form factors included in the vertices with baryon and meson resonances have
similar form to the bare $\pi N N$ form factor given in \eqref{pinn_bare},
with the half-width chosen the same for all the vertices, $\Lambda^2=1$ GeV$^2$.

The $\Delta N \pi$ and also the $\Delta N \gamma$ vertices have been
chosen so as to obey the gauge condition $p \cdot
\Gamma=0$~\cite{Pas98}. As a consequence of this the coupling to the
spin 1/2 components in the Rarita-Schwinger propagator are eliminated.
Only the vertices for the nucleon and the $\Delta$-resonance are given,
those for other $j=1/2$ and $j=3/2$ resonances are similar except for an
additional factor $\gamma_5$ for resonances carrying negative parity. A
cross in the propagator attached to a vertex implies that the particle
corresponding to this leg is taken on-shell.

The parameters $a_\pi$ and $a_\gamma$ appearing in the vertices for
spin--$3/2$ particles determine the ratio of coupling to positive-- and
negative--energy intermediate states. In principle the value of the
parameter $a_\gamma$ can be different for the two structures in the
photon-resonance vertex; however, these have been taken the same for
simplicity.

\setlength{\PicSize}{20mm}
\setlength{\DiagramWidth}{25mm}
\setlength{\FormulaWidth}{\textwidth}
\addtolength{\FormulaWidth}{-\DiagramWidth}
\addtolength{\FormulaWidth}{-5mm}

\vspace*{1ex} \noindent
%\begin{minipage}{\DiagramWidth}
%{\epsfxsize \PicSize \epsffile{../thesis/vrt_rho_pi_pi.ps} }
\begin{minipage}{\DiagramWidth}
{\epsfxsize \PicSize \epsffile{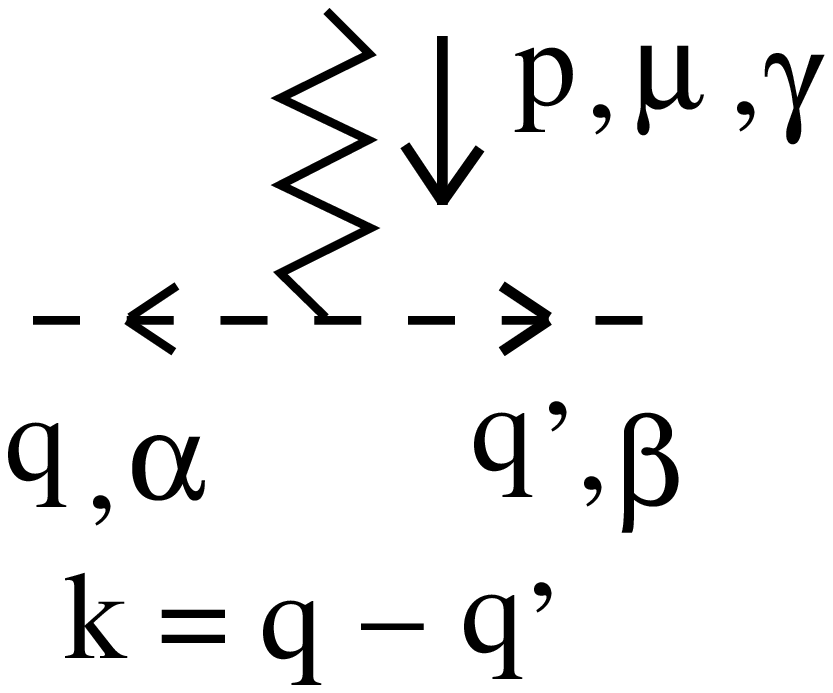} }
\end{minipage}
\begin{minipage}{\FormulaWidth}
$\displaystyle{
(\Gamma_{\rho \pi \pi})^{\,\mu}_{\,\alpha \beta \gamma} =
\epsilon_{\alpha \beta \gamma}\,g_{\rho \pi \pi} F_{\rho}(p^2)
\left[ k^{\mu}-
\frac{(p \cdot k)}{p^2}p^{\mu} \right]
} $ \end{minipage} 

\vspace*{1ex} \noindent
%\begin{minipage}{\DiagramWidth} {\epsfxsize \PicSize \epsffile{../thesis/vrt_sig_pi_pi.ps} }
\begin{minipage}{\DiagramWidth}
{\epsfxsize \PicSize \epsffile{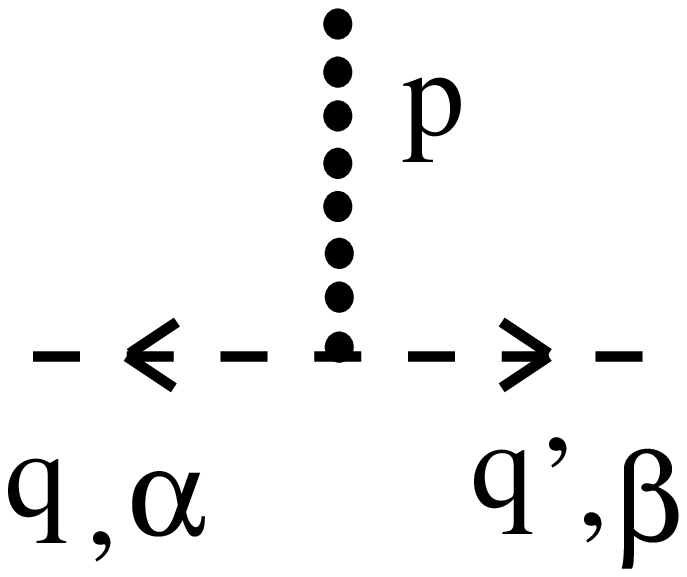} }
\end{minipage}
\begin{minipage}{\FormulaWidth}
$\displaystyle (\Gamma_{\sigma \pi \pi})_{\,\alpha \beta} =
-i\, F_{\sigma}(p^2) \delta_{\alpha \beta}\,
\left[ g_{\sigma \pi \pi}\, m_\pi  +
f_{\sigma \pi \pi}\, {(q\cdot q') \over m_\pi} \right]
$ \end{minipage}

\vspace*{1ex} \noindent
%\begin{minipage}{\DiagramWidth} {\epsfxsize \PicSize \epsffile{../thesis/vrt_rho_pi_gam.ps} }
\begin{minipage}{\DiagramWidth} {\epsfxsize \PicSize \epsffile{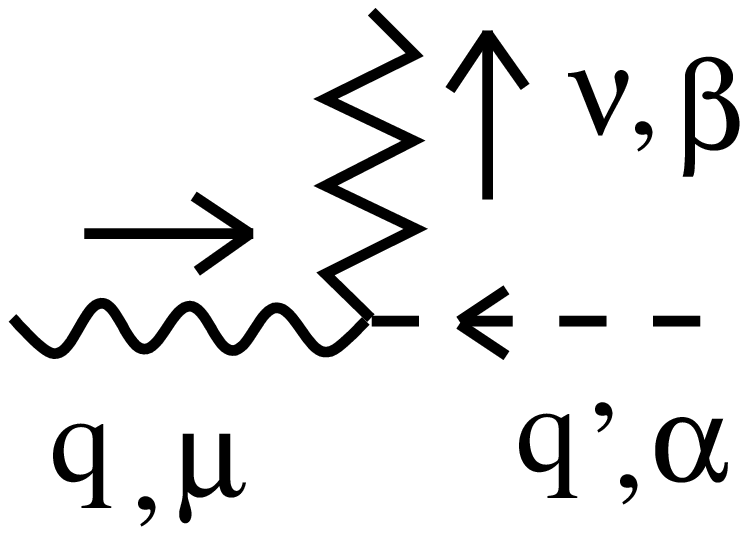} }
\end{minipage}
\begin{minipage}{\FormulaWidth}
$\ds  (\Gamma_{\rho \pi \gamma})^{\,\mu \nu}_{\alpha \beta}=
-i e \frac{g_{\rho \pi \gamma}}{m_\pi} \epsilon^{\mu \nu \rho \sigma}
q_\rho q^{\prime}_\sigma \, \delta_{\alpha \beta}
$ \end{minipage}

\vspace*{1ex} \noindent
%\begin{minipage}{\DiagramWidth} {\epsfxsize \PicSize \epsffile{../thesis/vrt_om_pi_gam.ps} }
\begin{minipage}{\DiagramWidth}
{\epsfxsize \PicSize \epsffile{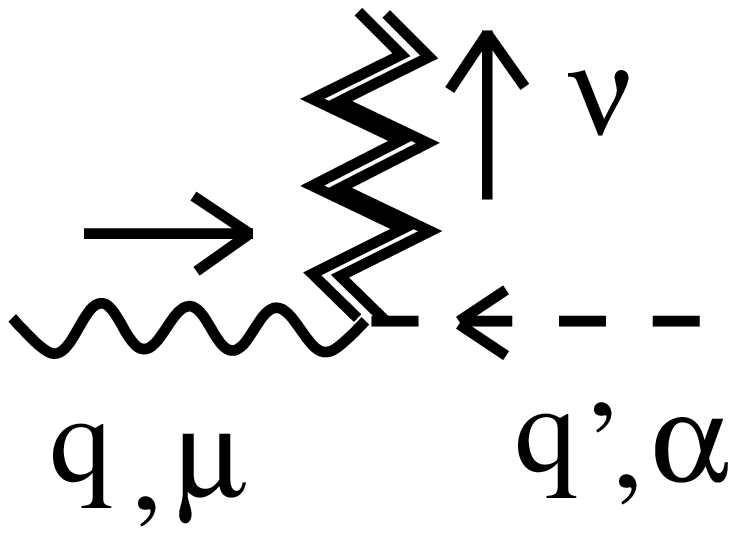} }
\end{minipage}
\begin{minipage}{\FormulaWidth}
$\ds  (\Gamma_{\omega \pi \gamma})^{\,\mu \nu}_\alpha =
-i e \frac{g_{\omega \pi \gamma}}{m_\pi} \epsilon^{\mu \nu \rho \sigma}
q_\rho q^{\prime}_\sigma \, \delta_{\alpha 3}
$ \end{minipage}

\vspace*{1ex} \noindent
%\begin{minipage}{\DiagramWidth} {\epsfxsize \PicSize \epsffile{../thesis/vrt_pi_gam_gam.ps} }
\begin{minipage}{\DiagramWidth} {\epsfxsize \PicSize \epsffile{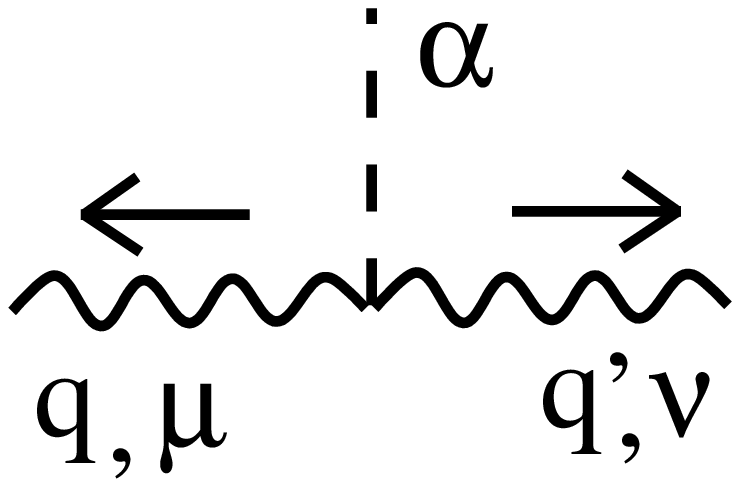} }
\end{minipage}
\begin{minipage}{\FormulaWidth}
$\ds (\Gamma_{\pi \gamma \gamma})^{\,\mu \nu}_{\,\alpha}=
-i \frac{e^2 g_{\pi \gamma \gamma}}{m_\pi} \epsilon^{\mu \nu \rho \sigma}
q_\rho q^{\prime}_\sigma \delta_{\alpha 3}
$ \end{minipage}

\vspace*{1ex} \noindent
%\begin{minipage}{\DiagramWidth} {\epsfxsize \PicSize \epsffile{../thesis/vrt_sig_gam_gam.ps} }
\begin{minipage}{\DiagramWidth} {\epsfxsize \PicSize \epsffile{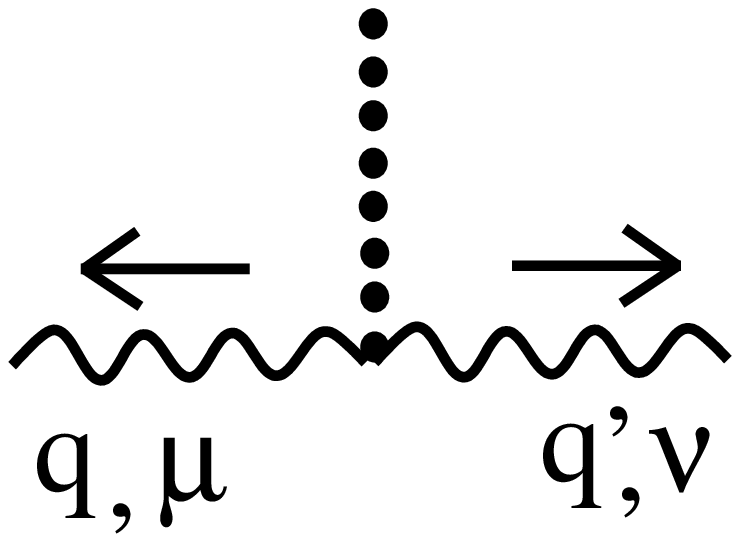} }
\end{minipage}
\begin{minipage}{\FormulaWidth}
$\ds  (\Gamma_{\sigma \gamma \gamma})^{\,\mu \nu}=
-i\frac{e^2 }{m_\sigma}
\left[\, q^{\prime \mu} \, q^\nu - \delta^{\mu \nu} \, (q \cdot q') \,\right]
\left[ g_{\sigma \gamma \gamma} + f_{\sigma \gamma \gamma}
         \frac{(q \cdot q')}{m_{\sigma}^2}  \right]
$ \end{minipage}

\newpage
\vspace*{1ex} \noindent
%\begin{minipage}{\DiagramWidth} {\epsfxsize \PicSize \epsffile{../thesis/vrt_pi_n_n.ps} }
\begin{minipage}{\DiagramWidth}
{\epsfxsize \PicSize \epsffile{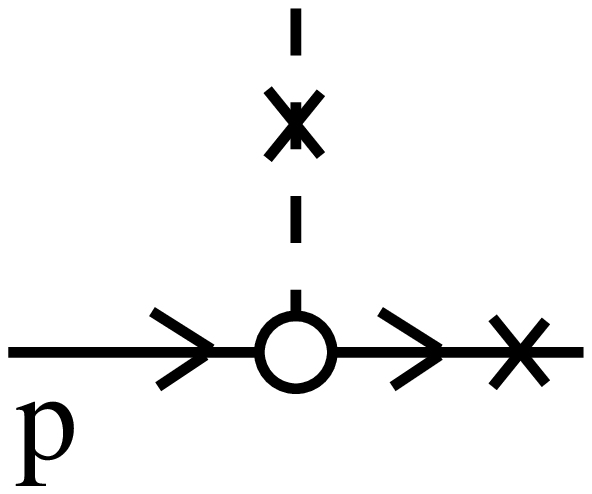} }
\end{minipage}
\begin{minipage}{\FormulaWidth}
$\displaystyle{
(\Gamma_{\pi N N})_\alpha =
\tau_\alpha\gamma^5
\left[ G_{ps}(p^2) + \frac{\vslash{p}+m}{2\, m} G_{pv}(p^2) \right]
} $ \end{minipage}

\vspace*{1ex} \noindent
%\begin{minipage}{\DiagramWidth} {\epsfxsize \PicSize \epsffile{../thesis/vrt_gam_n_n.ps} }
\begin{minipage}{\DiagramWidth}
{\epsfxsize \PicSize \epsffile{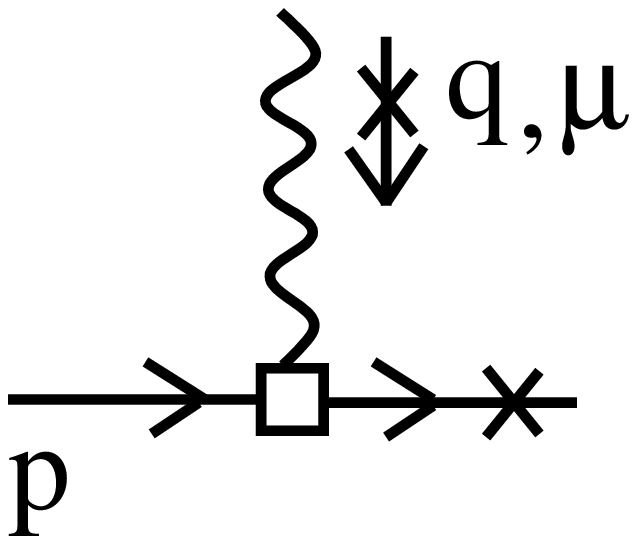} }
\end{minipage}
\begin{minipage}{\FormulaWidth}
$\displaystyle{
(\Gamma_{\gamma N N})^\mu = -ie
\sum_{l={\pm}}  \left[ \gamma^{\mu} \hat{F}_1^{l}(p^2) +
i \frac{\sigma^{\mu \nu}q_{\nu}}{2 m} \hat{F}_2^{l}(p^2)
\right]\frac{l \vslash{p}+m}{2 m}
}$
\end{minipage}

%%\newpage

\vspace*{1ex} \noindent
\begin{minipage}{\DiagramWidth}
%{\epsfxsize \PicSize \epsffile{../thesis/vrt_rho_n_n.ps} }
{\epsfxsize \PicSize \epsffile{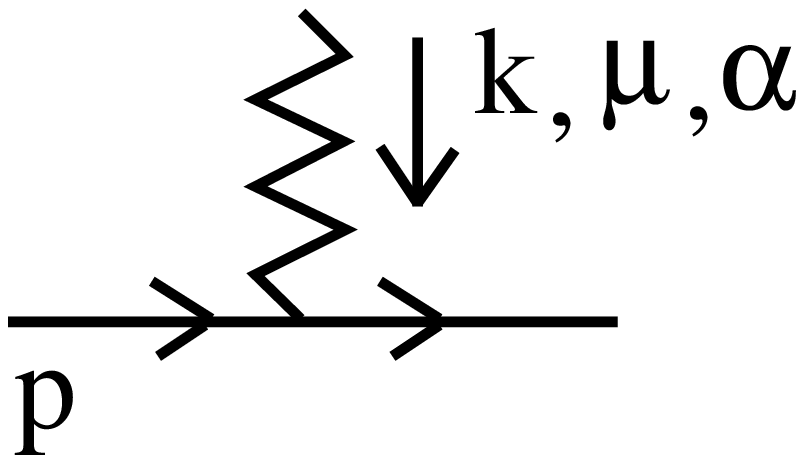} }
\end{minipage}
\begin{minipage}{\FormulaWidth}
$\displaystyle{
(\Gamma_{\rho N N})^{\,\mu}_{\,\alpha}
= -i\,g_{\rho N N} F_{N}(p^2) \frac{\tau_{\alpha}}{2}
\left( \gamma^\mu+i\,\kappa_\rho \frac{\sigma^{\mu \nu}k_\nu}{2m}
\right) } $ \end{minipage}

\vspace*{1ex} \noindent
\begin{minipage}{\DiagramWidth}
%{ \epsfxsize \PicSize \epsffile{../thesis/vrt_om_n_n.ps} }
{ \epsfxsize \PicSize \epsffile{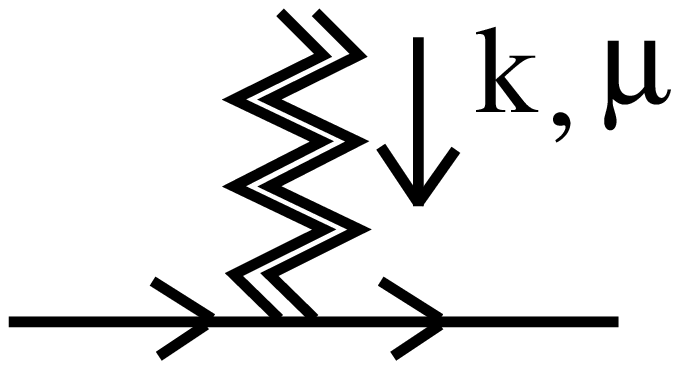} }
\hfill \end{minipage}
\begin{minipage}{\FormulaWidth}
$\displaystyle{ (\Gamma_{\omega N N})^{\,\mu}
= -i\,g_{\omega N N}
\left( \gamma^\mu+i\,\kappa_\omega \frac{\sigma^{\mu \nu}k_\nu}{2m}
\right) } $ \end{minipage}

\vspace*{1ex} \noindent
\begin{minipage}{\DiagramWidth}
%{\epsfxsize \PicSize \epsffile{../thesis/vrt_sig_n_n.ps} }
{\epsfxsize \PicSize \epsffile{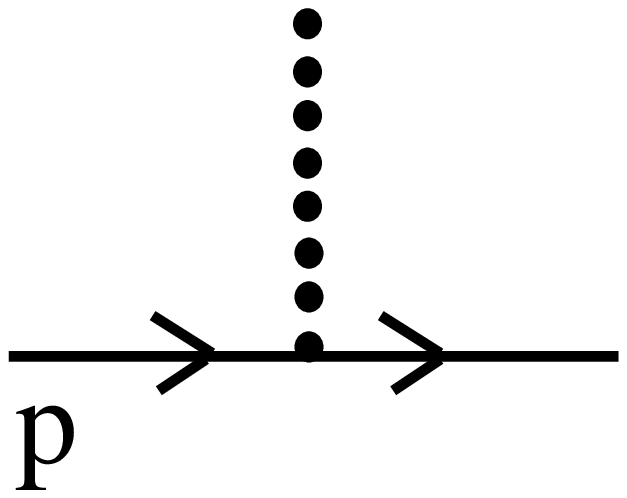} }
\end{minipage}
\begin{minipage}{\FormulaWidth}
$\displaystyle{
\Gamma_{\sigma N N}
= -i\,g_{\sigma N N} F_{N}(p^2)
} $ \end{minipage}

\vspace*{1ex} \noindent
%\begin{minipage}{\DiagramWidth} 
%{\epsfxsize \PicSize \epsffile{../thesis/vrt_del_n_pi.ps} }
\begin{minipage}{\DiagramWidth}
{\epsfxsize \PicSize \epsffile{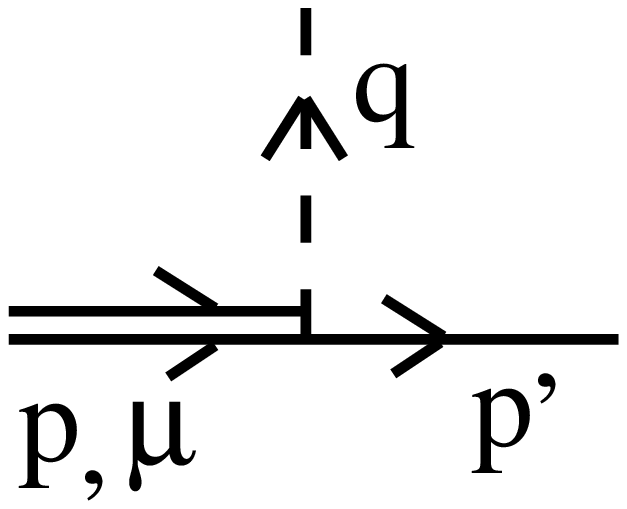} }
\end{minipage}
\begin{minipage}{\FormulaWidth}
$\displaystyle
(\Gamma_{\pi N \Delta})^{\,\mu}_{\,\alpha}
=i\,\frac{g_{\pi N \Delta}}{m_\pi^2}\, T_{\alpha}
 \, F_{\Delta}(p^2) \,  F_{N}(p^{\prime 2}) \,
 \left[ \,\vslash{p}q^{\mu}-(p \cdot q)\gamma^{\mu} \,\right]
\\ \hspace*{8em} % 1em=10pts for this font
\times \left[ (1-a_\pi)+a_\pi \frac{\vslash{p}}{m_\Delta} \right]
$ \end{minipage}

%\newpage
\vspace*{1ex} \noindent
%\begin{minipage}{\DiagramWidth} 
%{\epsfxsize \PicSize \epsffile{../thesis/vrt_del_n_gam.ps} }
\begin{minipage}{\DiagramWidth} {\epsfxsize \PicSize 
\epsffile{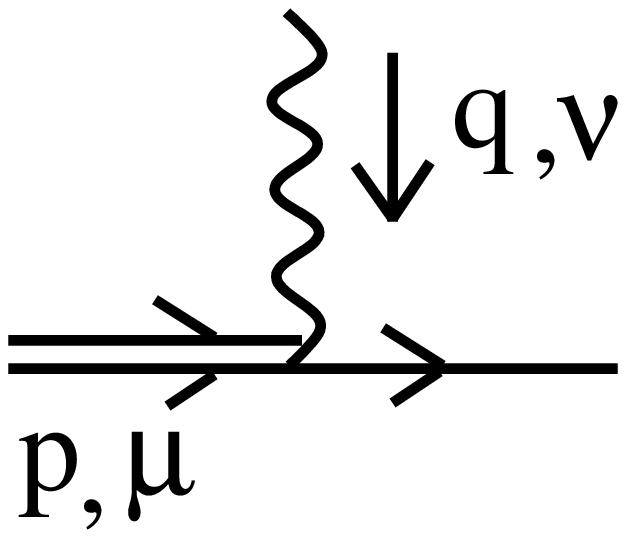} }
\end{minipage}
\begin{minipage}{\FormulaWidth}
$\ds
(\Gamma_{\Delta N \gamma})^{\,\mu \nu}=\hfill \\
\frac{i e}{2 m_\Delta^2}\Bigg\{\, g_1 \left[\, g^{\nu \mu}\vslash{p} 
\vslash{q}
-p^\nu \gamma^\mu \vslash{q}
-\gamma^\nu \gamma^\mu (p \cdot q) \right.
%\\ \hspace*{3.25cm}
\left. + \gamma^\nu q^\mu \vslash{p} \,\right] +
g_2 \left[\, q^\mu p^\nu - g^{\mu \nu}(p \cdot q)  \,\right]\Bigg\}
\\ \hspace*{8em} % 1em=10pts for this font
\times \left[ (1-a_\gamma)+a_\gamma \frac{\vslash{p}}{m_\Delta} \right]
\gamma^5\, T_3
$ \end{minipage}

\vspace*{1ex} \noindent
The normalization of the isospin $3/2$ to $1/2$ transition operators 
is chosen
according to 
$$ T_\alpha T_\beta^\dagger =\delta_{\alpha \beta} -
\frac{\tau_\alpha \tau_\beta}{3} $$ .

\begin{center}
\vspace*{-.5cm}
{\bf PROPAGATORS}
\end{center}

For cut propagators, the usual form is used as $2i$ multiplied by the imaginary
part of the propagator at positive energies; only the pole contribution is
taken into account in the case of the stable particles (nucleon and pion).

\setlength{\PicSize}{15mm}

\vspace*{1ex} \noindent
%\begin{minipage}{\DiagramWidth} {\epsfxsize \PicSize \epsffile{../thesis/prp_pi_free.ps} }
\begin{minipage}{\DiagramWidth} {\epsfxsize \PicSize \epsffile{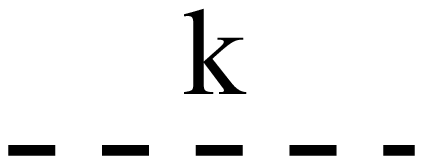} }
\end{minipage}
\begin{minipage}{\FormulaWidth}
$\displaystyle{
D_\pi^0=\frac{i}{k^2-m_\pi^2+i0}
} $ \end{minipage}

\newpage
\vspace*{1ex} \noindent
%\begin{minipage}[t]{\DiagramWidth} {\epsfxsize \PicSize \epsffile{../thesis/prp_sig_dres.ps} }
\begin{minipage}[t]{\DiagramWidth}
{\epsfxsize \PicSize \epsffile{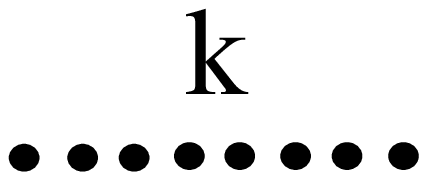} }
\end{minipage}
\begin{minipage}[t]{\FormulaWidth}
$\displaystyle D_\sigma= \frac{i}{k^2-m_\sigma^2-\Pi_\sigma(k^2)},
$ \\[2ex] \hspace*{0.9cm} $\displaystyle
\Pi_\sigma(k^2)=\Pi_{\sigma, L}(k^2)-(Z_\sigma-1)(k^2-m_\sigma^2)-
Z_\sigma \delta m_\sigma^2
$ \end{minipage}

\vspace*{1ex} \noindent
%\begin{minipage}[t]{\DiagramWidth} {\epsfxsize \PicSize \epsffile{../thesis/prp_rho_dres.ps} }
\begin{minipage}[t]{\DiagramWidth}
{\epsfxsize \PicSize \epsffile{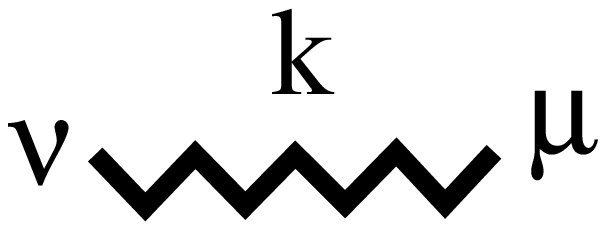} }
\end{minipage}
\begin{minipage}[t]{\FormulaWidth}
$\displaystyle
(D_\rho)^{\mu \nu}=
\frac{-i\,{\mathcal{P}}_1^{\mu \nu}(k)}{k^2-m_\rho^2-\Pi_\rho(k^2)}, \;
{\mathcal{P}}_1^{\mu \nu}(k)=g^{\mu \nu}-\frac{k^\mu k^\nu}{k^2},
$ \\[2ex] \hspace*{1.5cm} $\displaystyle
\Pi_\rho(k^2)=\Pi_{\rho, L}(k^2)-(Z_\rho-1)(k^2-m_\rho^2)-
Z_\rho \delta m_\rho^2
$ \end{minipage}

\vspace*{1ex} \noindent
%\begin{minipage}{\DiagramWidth} {\epsfxsize \PicSize \epsffile{../thesis/prp_om_free.ps} }
\begin{minipage}{\DiagramWidth}
{\epsfxsize \PicSize \epsffile{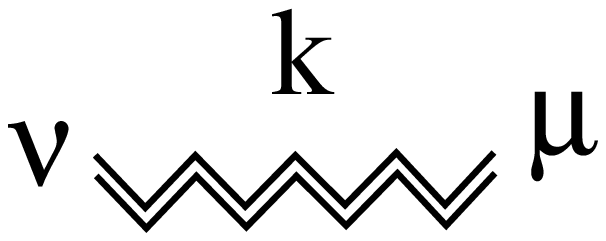} }
\end{minipage}
\begin{minipage}{\FormulaWidth}
$\displaystyle (D_\omega^0)^{\mu \nu}=
\frac{-i\,{\mathcal{P}}_1^{\mu \nu}(k)}{k^2-m_\omega^2+i0} $
\end{minipage}

\vspace*{1ex} \noindent
%\begin{minipage}[t]{\DiagramWidth} {\epsfxsize \PicSize \epsffile{../thesis/prp_n_dres.ps} }
\begin{minipage}[t]{\DiagramWidth}
{\epsfxsize \PicSize \epsffile{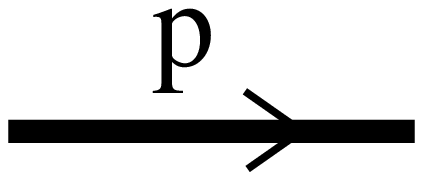} }
\end{minipage}
\begin{minipage}[t]{\FormulaWidth}
$\displaystyle{
S_N=\frac{i}{\vslash{p}-m-\Sigma_N(p)+i0}},\; $ \\[2ex]
\hspace*{0.8cm}
$\ds{ \Sigma_N(p)=
%\stackrel{\displaystyle{\Sigma_{N,L}(p)}}{\overbrace{
A_N(p^2)\vslash{p}+B_N(p^2)m
%}}}$ \\[2ex] \hspace*{1.9cm} $\displaystyle{
-(Z_2^N-1)(\vslash{p}-m) -Z_2^N \delta m
} $ \end{minipage}

\vspace*{1ex} \noindent
%\begin{minipage}[t]{\DiagramWidth} {\epsfxsize \PicSize
%\epsffile{../thesis/prp_del_dres.ps} }
\begin{minipage}[t]{\DiagramWidth} {\epsfxsize \PicSize
\epsffile{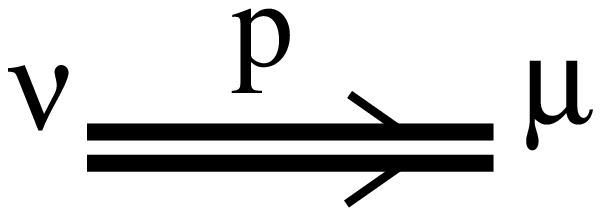} }
\end{minipage}
\begin{minipage}[t]{\FormulaWidth}
$\displaystyle{
(S_\Delta)^{\mu \nu}=\frac{-i}
{\vslash{p}-m_\Delta-\Sigma_\Delta(p)} {\mathcal{P}}_{3/2}^{\mu \nu}(p)},\;$\\[2ex]
\hspace*{1.4cm}
$\ds \Sigma_\Delta(p)=
%\stackrel{\displaystyle{\Sigma_{\Delta,L}(p)}} {\overbrace{
A_\Delta(p^2)\vslash{p}+B_\Delta(p^2)m_\Delta
%}} $ \\[2ex] \hspace*{2.6cm} $\displaystyle
-(Z_2^\Delta-1)(\vslash{p}-m_\Delta) %-Z_2^\Delta \delta m_\Delta,
$ \\[2ex] \hspace*{1.4cm} $\displaystyle
{\mathcal{P}}_{3/2}^{\mu \nu}(p)=g^{\mu \nu}
-\frac{1}{3}\gamma^\mu \gamma^\nu
-\frac{1}{3p^2}(\vslash{p}\gamma^\mu p^\nu + p^\mu \gamma^\nu \vslash{p})
$ \end{minipage}

\vspace*{1ex} \noindent
Due to the gauge-invariance condition imposed on the spin-$3/2$
vertices, the terms in the $\Delta$ propagator  
proportional to $p^\mu$ and $p^\nu$
give vanishing contributions to the matrix elements.

\begin{table}[!htb]
\caption[t1]{Meson-nucleon and meson-meson coupling parameters of the 
model which were optimized to reproduce 
$\pi N$ scattering phase shifts, pion photoproduction and
Compton scattering on the proton. The parameter
$\Lambda^2_N$ is given in GeV$^{\,2}$.}
\begin{center}
\begin{tabular}{c|cc|cc|ccccc}
 $\Lambda^2_N$ & $g_{\rho N N}$ & $\kappa_\rho$ &
$g_{\omega N N}$  & $\kappa_\omega$ &
$g_{\sigma N N}$ &  $g_{\sigma \pi \pi}$ &  $f_{\sigma \pi \pi}$ &
 $g_{\sigma \gamma \gamma}$ &  $f_{\sigma \gamma \gamma}$   \\
\hline
 1.8 & 7.0 & 2.3 &
  12 & -0.8 &
  34 & 1.7 & 1.8 &
  -0.42 & -1.7  \\
\end{tabular}
\end{center}
\tablab{mesons}
\end{table}

\begin{table}[!htb]
\begin{center}
\caption[t2]{Parameters for the different $N^*$ resonances: masses (MeV), 
         one-pion couplings, two- (and multi-)pion widths (MeV),
         and off-shell parameters.}
\begin{tabular}{c|rr|rr|rrr}
  $N^*$ resonance & $M_r$ & $\Gamma_0$ & $g_{\pi N^*N}$ & $a_\pi$ &
  $g_{1,\gamma N^*N}$ & $g_{2,\gamma N^*N}$ & $a_{\gamma}$ \\
\hline
   $P_{11}(1440)$ & 1550 &  80 & 11.1  & 0.08 &  1.2 & -- & -- \\
   $D_{13}(1520)$ & 1500 &  90 &  1.3  & 0.35 &  4.5 & 5.9 & 0.65 \\
   $S_{11}(1535)$ & 1540 &  80 &  1.8  & 1.05 &  -1.5 & -- & -- \\
   $S_{11}(1650)$ & 1720 & 100 &  3.9  & 1.05 &  -2.2 & -- & --\\
   $P_{13}(1710)$ & 1720 & 220 &  0.22 & 0.5  &   0   & -3.0 & 0.5\\
   $P_{33}(1232)$ & 1230 &   0 &  2.48 & 0    & -2.33 & -3.02 &  -2.40 \\
   $S_{31}(1620)$ & 1600 &  30 &  2.25 & 0.75 & -0.20 & -- & -- \\
   $D_{33}(1700)$ & 1650 & 300 &  0.37 & 0.5  &  1.70 & 0  & 0.5  \\
\hline
\end{tabular}
\tablab{reson}
\end{center}
\end{table}

\begin{table}[!htb]
\caption[T4]{Polarizabilities of the proton. The units are
$10^{-4}fm^3$ for $\alpha$ and $\beta$ and $10^{-4}fm^4$ for the
$\gamma$s. The $\gamma$s are given without the anomalous
$\pi^0$ contribution.
The first two columns contain the polarizabilities obtained from the
present calculation (dressed and bare).
The three columns named $\chi PT$ contain the
polarizabilities calculated in the chiral perturbation theory:
leading order, next-to-leading order and ${\mathcal{O}}(\epsilon^3)$,
from left to right.
Results of recent dispersion analyses are given in the last column
(Ref.~\cite{Lvo97} for $\alpha$ and $\beta$ and Ref.~\cite{Dre00} for the
$\gamma$s). } 
\begin{center}
\begin{tabular}{c|cc|ccc|c}
 & D & B & & $\chi$PT &  &   DA   \\
 &   &   & \cite{Ber95} & \cite{Ber_K93,Gel00} $(\;\mbox{\cite{Vij00}}\;)$
 & \cite{Hem98} &   \cite{Lvo97,Dre00}     \\
\hline
$\alpha$    &12.1 & 15.5 &12.2 &10.5$\pm$2.0 &16.4              &11.9   \\
$\beta$     & 2.4 & 1.7  & 1.2 & 3.5$\pm$3.6 & 9.1              & 1.9   \\
$\gamma_{E1}$ &-5.0 & -1.7 &-5.7 &-1.9 $(\;\mbox{-}1.3\;)$ &-5.4  & -4.3  \\
$\gamma_{M1}$ & 3.4 & 3.8  &-1.1 & 0.4 $(\;3.3\;)$ & 1.4          & 2.9   \\
$\gamma_{E2}$ & 1.1 & 1.0  & 1.1 & 1.9 $(\;1.8\;)$ & 1.0          & 2.2   \\
$\gamma_{M2}$ &-1.8 & -2.3 & 1.1 & 0.7 $(\;0.2\;)$ & 1.0          & 0.0   \\
$\gamma_0$    & 2.4 & -0.9 & 4.6 &-1.1 $(\;\mbox{-}4.0\;)$ & 2.0 &-0.8   \\
$\gamma_\pi$  & 11.4& 8.9  & 4.6 & 3.5 $(\;6.2\;)$ & 6.8         & 9.4   \\
\hline
\end{tabular}
\end{center}
\tablab{polp}
\end{table}

\begin{table}[!htb]
\caption[T5]{Polarizabilities of the neutron. Explanation of the entries is as
in \tabref{polp}.}
\begin{center}
\begin{tabular}{c|cc|ccc|c}
 & D & B & & $\chi$PT &  &   DA    \\
 &   &   & \cite{Ber95} & \cite{Ber_K93,Gel00} $(\;\mbox{\cite{Vij00}}\;)$
 & \cite{Hem98} &  \cite{Lvo97,Dre00}               \\
\hline
$\alpha$    & 12.7 & 15.7 &12.2 &13.4$\pm$1.5    &16.4 &13.3  \\
$\beta$     & 1.8  & 1.4  & 1.2 & 7.8$\pm$3.6     & 9.1 & 1.8  \\
$\gamma_{E1}$ &-4.8  &-1.7  &-5.7 &-4.3 $(\;4.0\;)$ &-5.4 & -6.0  \\
$\gamma_{M1}$ & 3.5  & 4.1  &-1.1 & 0.4 $(\;2.3\;)$ & 1.4 &  3.9    \\
$\gamma_{E2}$ & 1.1  & 1.0  & 1.1 & 2.3 $(\;2.2\;)$ & 1.0 & 3.2  \\
$\gamma_{M2}$ & -1.8 & -2.3 & 1.1 & 0.5 $(\;0.4\;)$ & 1.0 &-1.0   \\
$\gamma_0$    & 2.0  & -1.1 & 4.6 & 1.1 $(\;\mbox{-}0.9\;)$ & 2.0 &-0.1  \\
$\gamma_\pi$  & 11.2 & 9.1  & 4.6 & 6.5 $(\;8.1\;)$ & 6.8 &14.1 \\
\hline
\end{tabular}
\end{center}
\tablab{poln}
\end{table}

\begin{table}[!htb]
\caption[T6]{The various contributions to the proton polarizabilities are 
given. The notation is explained in the text.}
\begin{center}
\begin{tabular}{r|r|r|c|c|c|l|r|r}
 & $\alpha$ & $\beta$ &  $\gamma_{E1}$ & $\gamma_{M1}$ &
 $\gamma_{E2}$ & $\gamma_{M2}$ & $\gamma_0$ & $\gamma_\pi$  \\
\hline
 Full        & 12.1 &   2.4 & -5.0 & 3.4 & 1.1 & -1.8 & 2.4 & 11.4 \\
no $\Delta$  & 13.9 & -11.2 & -3.7 & 0.8 & 0.4 & -0.07& 2.6 & 4.9 \\
no $\sigma$  &  1.3 &  13.2 & -5.0 & 3.4 & 1.1 & -1.8 & 2.4 & 11.4 \\
no cusp      &  8.9 &   2.4 & -1.7 & 3.1 & 0.8 & -1.8 &-0.3 & 7.4 \\
\hline
\end{tabular}
\end{center}
\tablab{polp-contr}
\end{table}  

\begin{table}[!htb]
\caption[T7]{ Same as in \tabref{polp-contr}, but for the neutron polarizabilities.}
\begin{center}
\begin{tabular}{r|r|r|c|c|c|l|r|r}
 & $\alpha$ & $\beta$ &  $\gamma_{E1}$ & $\gamma_{M1}$ &
 $\gamma_{E2}$ & $\gamma_{M2}$ & $\gamma_0$ & $\gamma_\pi$  \\
\hline
 Full        & 12.7 &   1.8 & -4.8 & 3.5 & 1.1 & -1.8 & 2.0 & 11.2 \\
no $\Delta$  & 14.6 & -11.8 & -3.5 & 0.9 & 0.3 & -0.08& 2.3 &  4.8 \\
no $\sigma$  &  1.9 &  12.6 & -4.8 & 3.5 & 1.1 & -1.8 & 2.0 & 11.2 \\
no cusp      &  9.6 &   1.8 & -1.5 & 3.2 & 0.7 & -1.9 &-0.6 &  7.2 \\
\hline
\end{tabular}
\end{center}
\tablab{poln-contr}
\end{table}

\begin{figure}[!htb]
%\centerline{{\epsfxsize 9cm \epsffile{eq_piNN_1.ps}}}
\centerline{{\epsfxsize 9.cm \epsffile[0 170 590 680]{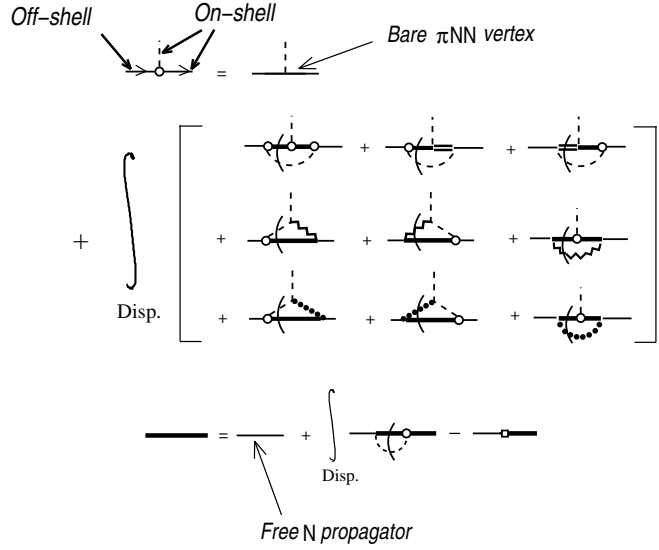}}}
\caption[f1]{Graphical representation of the equation for the dressed
irreducible $\pi N N$ vertex, denoted by an open circle, and the dressed
nucleon propagator, denoted by a solid line.
The dashed lines denote pions,
the double lines denote $\Delta$s and
the zigzag and dotted lines are $\rho$ and $\sigma$ mesons, respectively.
The resonance propagators are dressed. 
The last term in the second equation denotes the counter-term contribution
to the nucleon propagator, necessary for the renormalization.
\figlab{piNN}}
\end{figure}

\begin{figure}[!htb]
%\centerline{{\epsfxsize 10cm \epsffile{..thesis/eq_gamNN.ps}}}
\centerline{{\epsfxsize 10.cm \epsffile{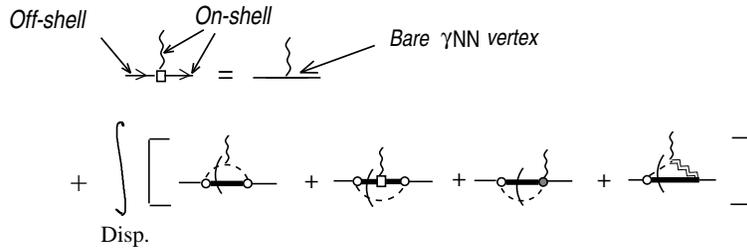}}}
\caption[f2]{ {Equation for the irreducible $\gamma N N$ vertex
(denoted by the square) used in the full model. The wiggly lines are photons,
the double zigzag line
is an $\omega$ meson. See \figref{piNN} for an explanation of the other
notation. }
\figlab{gamNN}} 
\end{figure}

\newpage
\begin{figure}[!htb]
%\centerline{\epsfxsize 8cm {\epsffile{..thesis/kmat_pipi.ps}}}
\centerline{\epsfxsize 8.cm {\epsffile{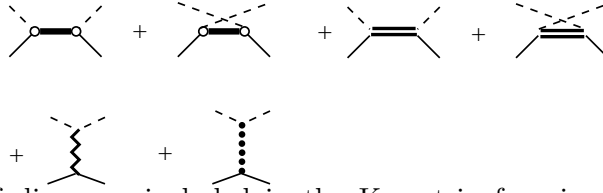}}}
\caption[f3]{The sum of diagrams included in the K matrix for
pion-nucleon scattering. The notation is explained in \figref{piNN}
\figlab{K-pipi}. The full spectrum of baryon resonances given in
\tabref{reson} have been included.}
\end{figure}

\begin{figure}[!htb]
%\centerline{\epsfxsize 8cm {\epsffile{../thesis/kmat_pig.ps}} }
\centerline{\epsfxsize 8cm {\epsffile{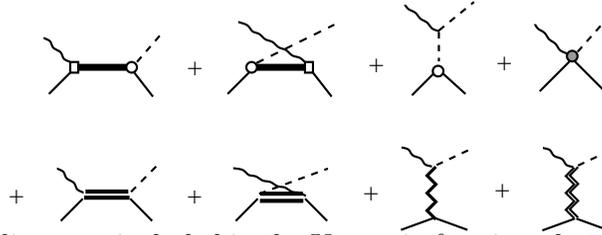}} }
\caption[f4]{The sum of diagrams included in the K matrix for
pion photoproduction. The notation is as in \figref{piNN} and \figref{gamNN}
The shaded
circle is the contact $\gamma \pi N N$ vertex. \figlab{K-gampi}}
\end{figure}

\begin{figure}[!htb]
%\centerline{{\epsfxsize 8cm \epsffile{../thesis/kmat_gg.ps}}}
\centerline{{\epsfxsize 8cm \epsffile{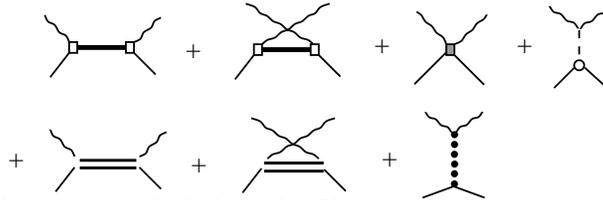}}}
\caption[f5]{The sum of diagrams included in the K matrix for
Compton scattering. The notation is as in \figref{K-pipi} and
\figref{K-gampi}.
The shaded square is the contact $\gamma \gamma N N$ vertex. \figlab{K-gg}}
\end{figure}

\newpage
\begin{figure}[!htb]
%\centerline{ \epsfxsize 6cm \epsffile{../figures/fit-b/RE_VS.PS}}
\centerline{ \epsfxsize 6cm \epsffile{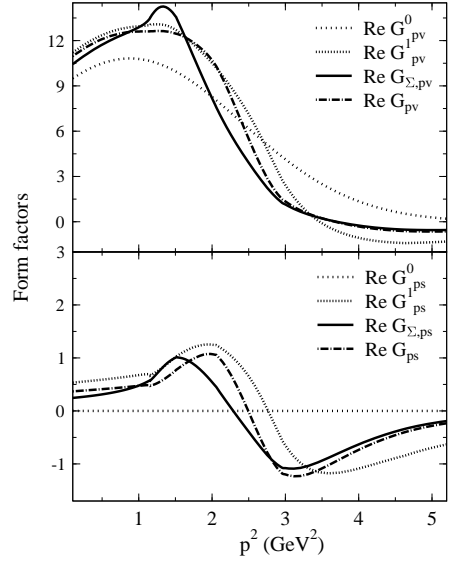}}
\caption[f6]{Dependence of the form factors $G_{pv}$ and $G_{ps}$
entering in the $\pi N N$ vertex, on
the momentum squared of the off-shell nucleon. Curves labelled with
superscript $^0$ ($^1$) show the bare form factor (results of the first
iteration). The converged form factors are given in two
different representations:  one where the nucleon self-energy is non-trivial 
($G_\Sigma$)
and one where the self-energy has been transformed out ($G$).
\figlab{f-pi}}    
\end{figure}

\begin{figure}[!htb]
%\centerline{ \epsfxsize 5cm \epsffile{../figures/fit-b/re_ab.ps}}
\centerline{ \epsfxsize 5cm \epsffile{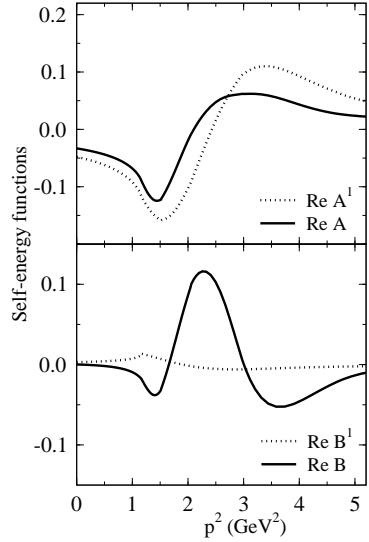}}
\caption[f7]{Dependence of the self-energy functions $A$ and $B$, as
enter in the nucleon self-energy, on
the momentum squared of the proton. Fully converged results and those
from the first iteration are given. \figlab{n-self}}
\end{figure}

\begin{figure}[!htb]
%\centerline{ \epsfxsize 6cm \epsffile{../figures/fit-b/f1.ps}}
\centerline{ \epsfxsize 6cm \epsffile{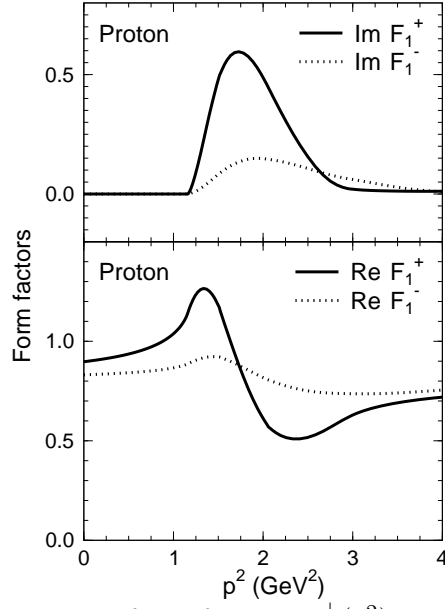}}
\caption[f8]{Dependence of the electric form factors $F_1^{\pm}(p^2)$ on
the momentum squared of the off-shell proton.
\figlab{f1-p}}
\end{figure} 

\begin{figure}[!htb]
\centerline{\begin{minipage}{6cm}
%{ \epsfxsize 5.5cm \epsffile[0 0 415 570]{../figures/fit-b/f2_pr.ps} }
{ \epsfxsize 5.5cm \epsffile[0 0 415 570]{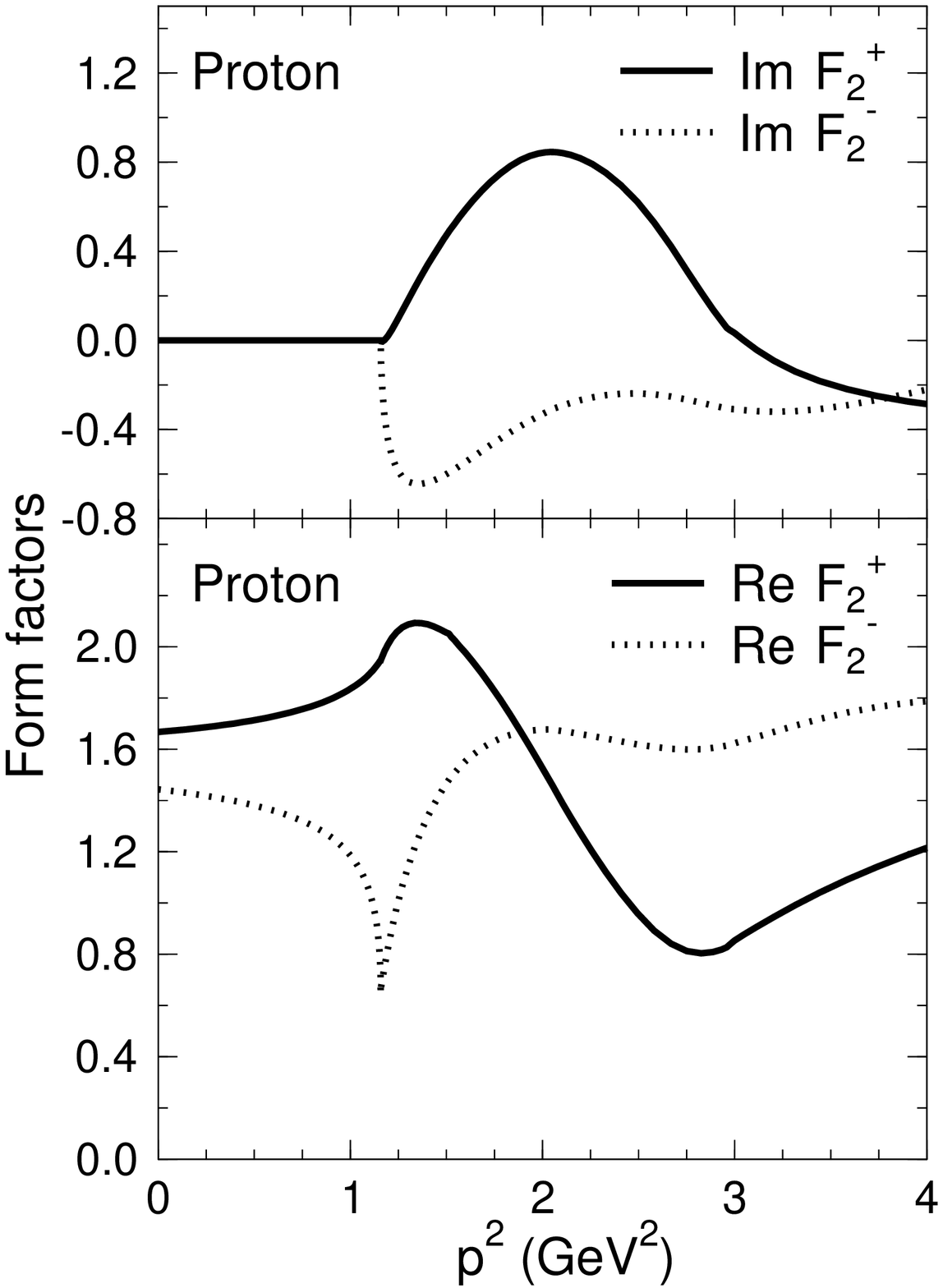} }
\end{minipage}
\begin{minipage}{6cm}
%{ \epsfxsize 5.5cm \epsffile[0 0 415 570]{../figures/fit-b/f2_ne.ps} }
{ \epsfxsize 5.5cm \epsffile[0 0 415 570]{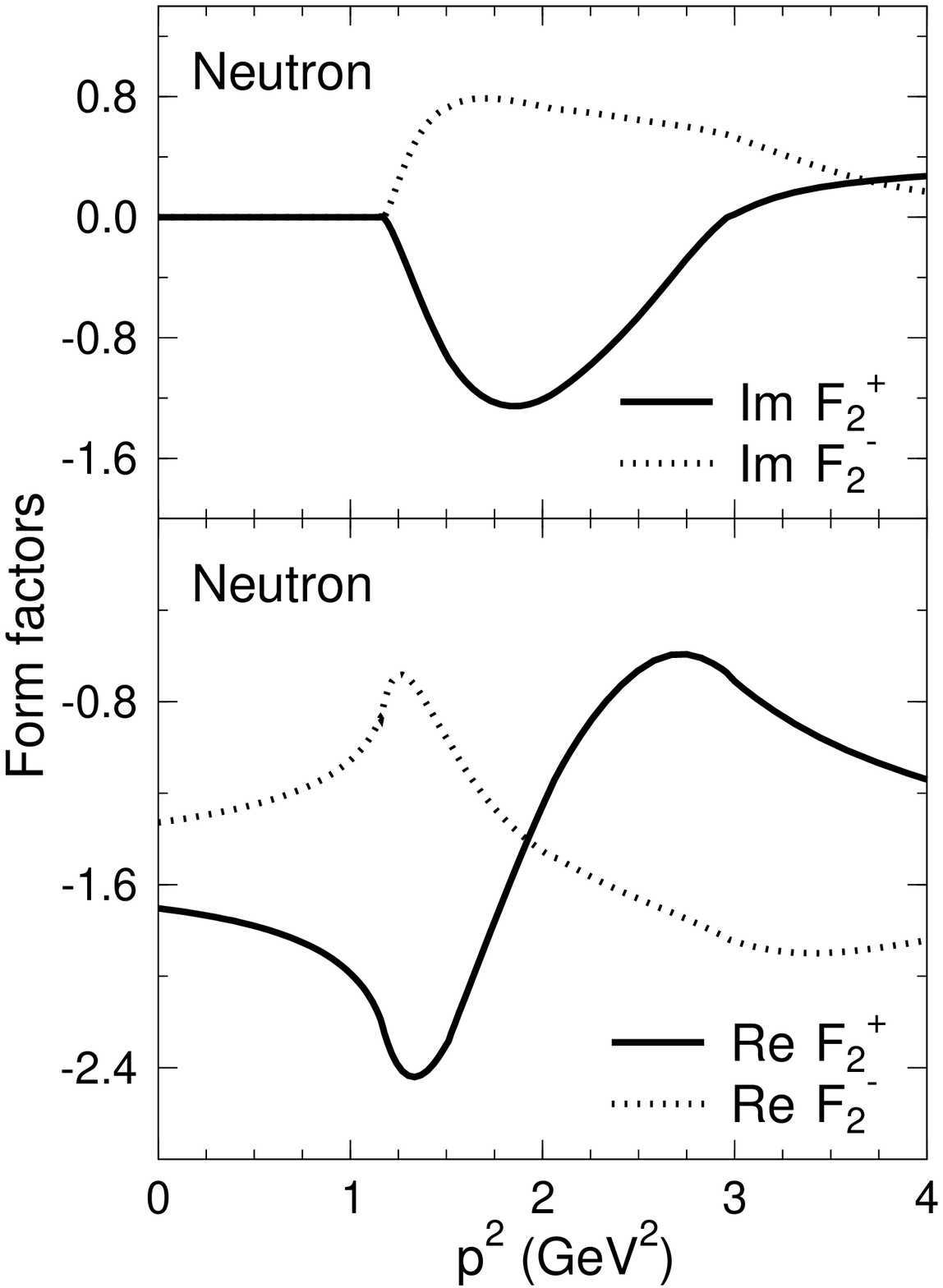} }
\end{minipage}}
\caption[f9]{Magnetic half-off-shell form factors $F_2^{\pm}(p^2)$
for the proton (left panel) and the neutron (right panel).
\figlab{f2}}
\end{figure}

\begin{figure}[!htb]
\centerline{ \epsfxsize 13.5cm
%\rotate [r]{\epsffile[0 0 550 700]{../figures/fit-b/pi-pi.ps}}}
\rotate [r]{\epsffile[0 0 550 700]{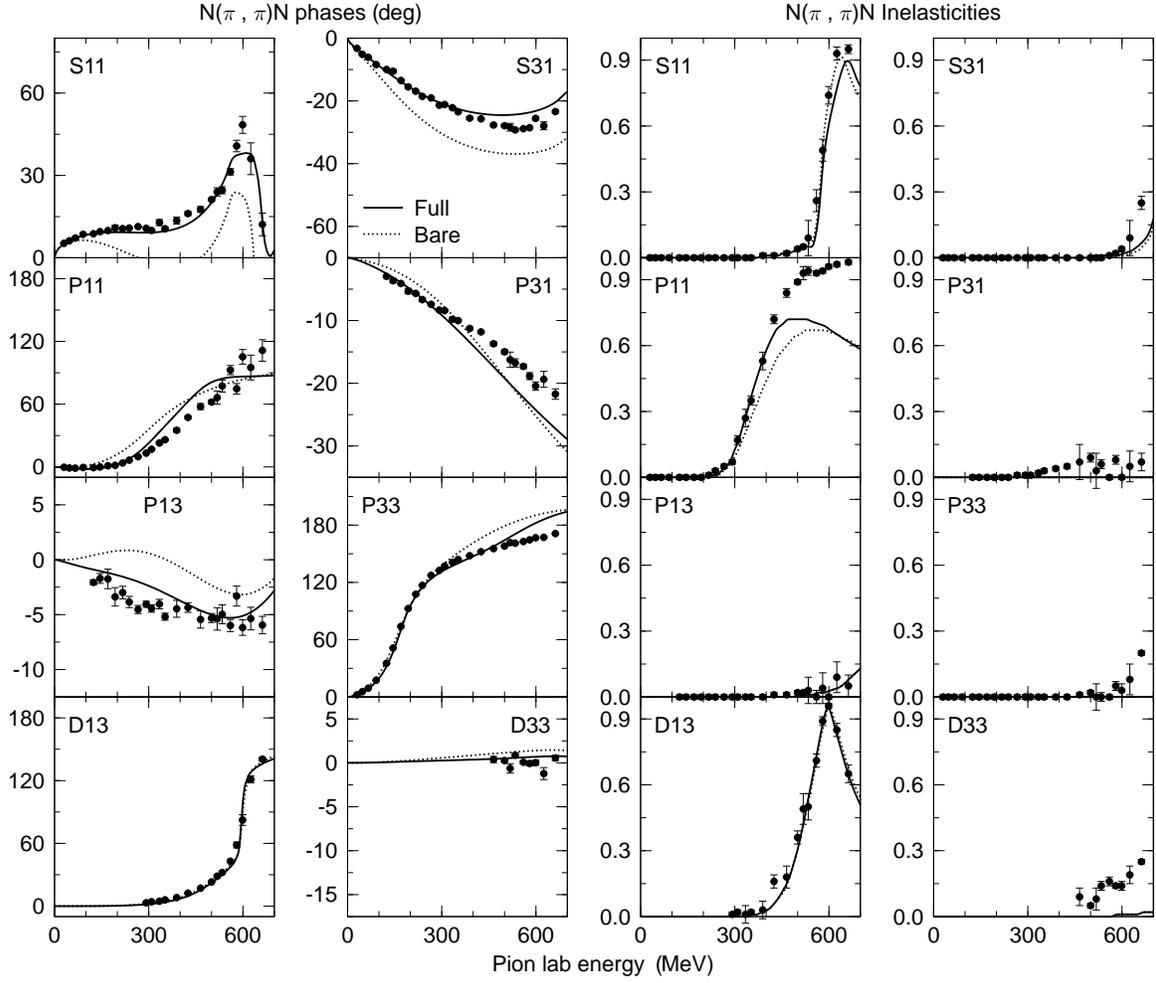}}}
\caption[f10]{Phase shifts and inelasticities for pion-nucleon
scattering from the analysis of Ref.~\cite{Arn95} are compared with the
results of the calculations  D (full lines) and B (dotted lines). \figlab{pi-pi}}
\end{figure}

\begin{figure}[!htb]
\centerline{ \epsfxsize 13.5cm
%\rotate [r]{\epsffile[0 0 550 700]{../figures/fit-b/gp_pw.ps}}}
\rotate [r]{\epsffile[0 0 550 700]{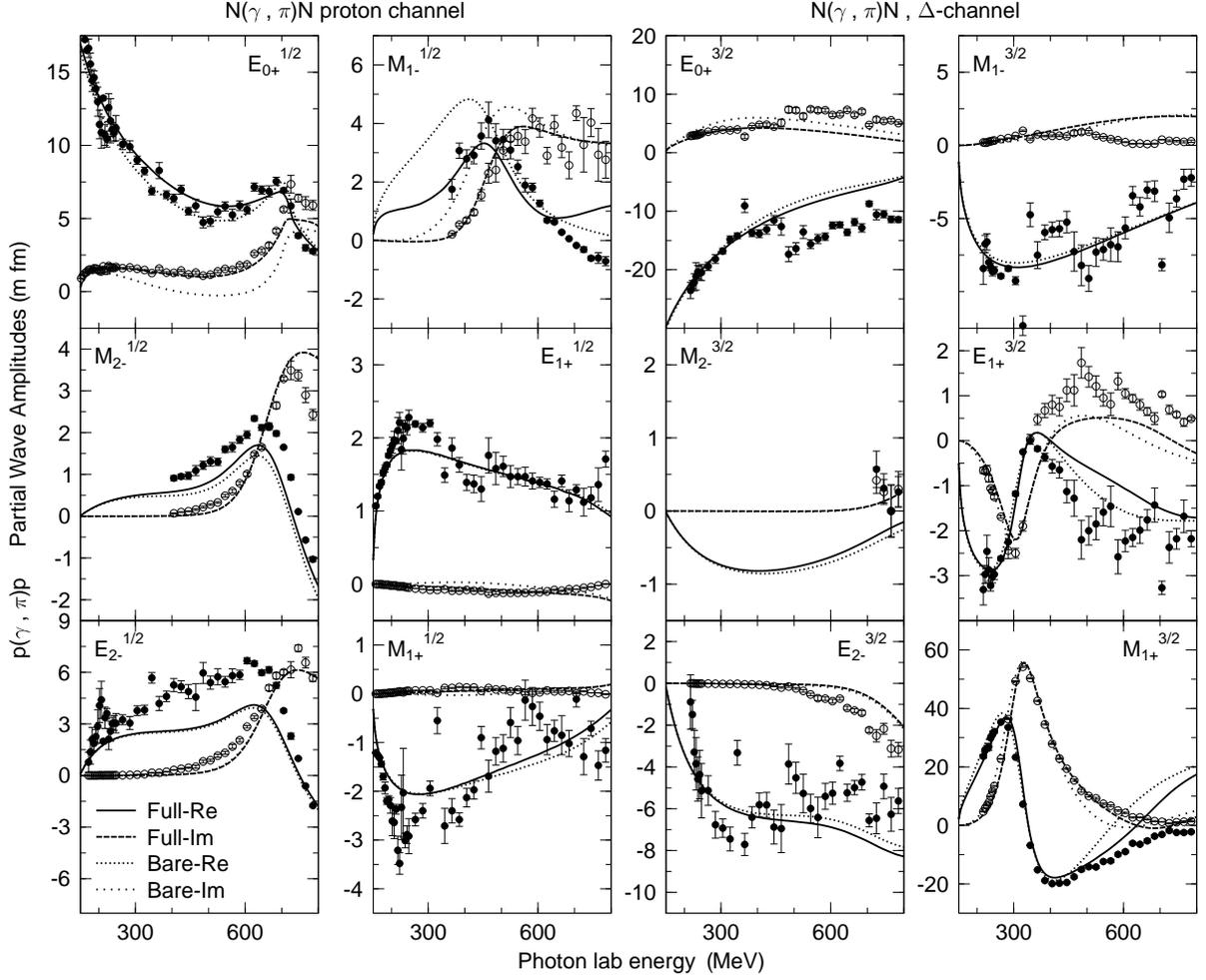}}}
\caption[f11]{
The calculated pion photoproduction
multipoles are compared to the partial-wave
analysis results from Ref.~\cite{Arn96}.
The solid and dashed lines are the real and imaginary parts,
respectively, of the multipoles from calculation D.
The dense-- and sparse-dotted lines are the corresponding
quantities obtained without the dressing, i.e.\ from calculation B.
\figlab{pi-gam}}
\end{figure}

%\begin{figure}[!htb]
%%\centerline{ \epsfxsize 9cm \epsffile{../figures/fit-b/ggs.ps}}
%\centerline{ \epsfxsize 9cm \epsffile{ggs.ps}}
%\caption[f12]{
%The angular distributions for Compton scattering are compared with the
%calculation at different energies.  Also shown are
%the energy distributions at the scattering angles of 75$^o$ and 90$^o$.
%The solid and dotted lines are from the calculations D and B, respectively.
%The data points are taken from \cite{Hal93} for the angular distributions and
%from \cite{Hun97} for the energy distributions.
%\figlab{ggs}}
%\end{figure}
\begin{figure}[!htb]
%\centerline{ \epsfxsize 9cm \epsffile{../figures/fit-b/ggs.ps}}
\centerline{ \epsfxsize 9cm \epsffile[0 270 480 700]{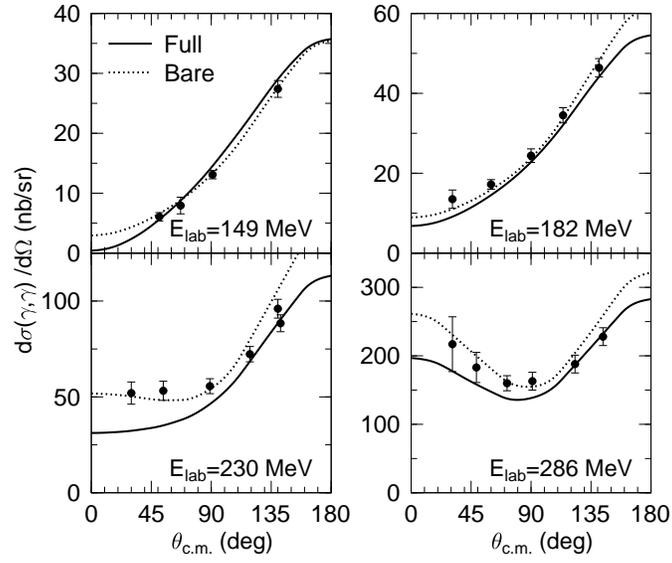}}
\caption[f12]{
The angular distributions for Compton scattering are compared with the
calculation at different energies.  
The solid and dotted lines are from the calculations D and B, respectively.
The data points are taken from \cite{Hal93}.
\figlab{ggs}}
\end{figure}

\begin{figure}[!htb]
%\centerline{ \epsfxsize 6cm \epsffile{../figures/fit-b/gam.ps}}
\centerline{ \epsfxsize 10 cm \epsffile{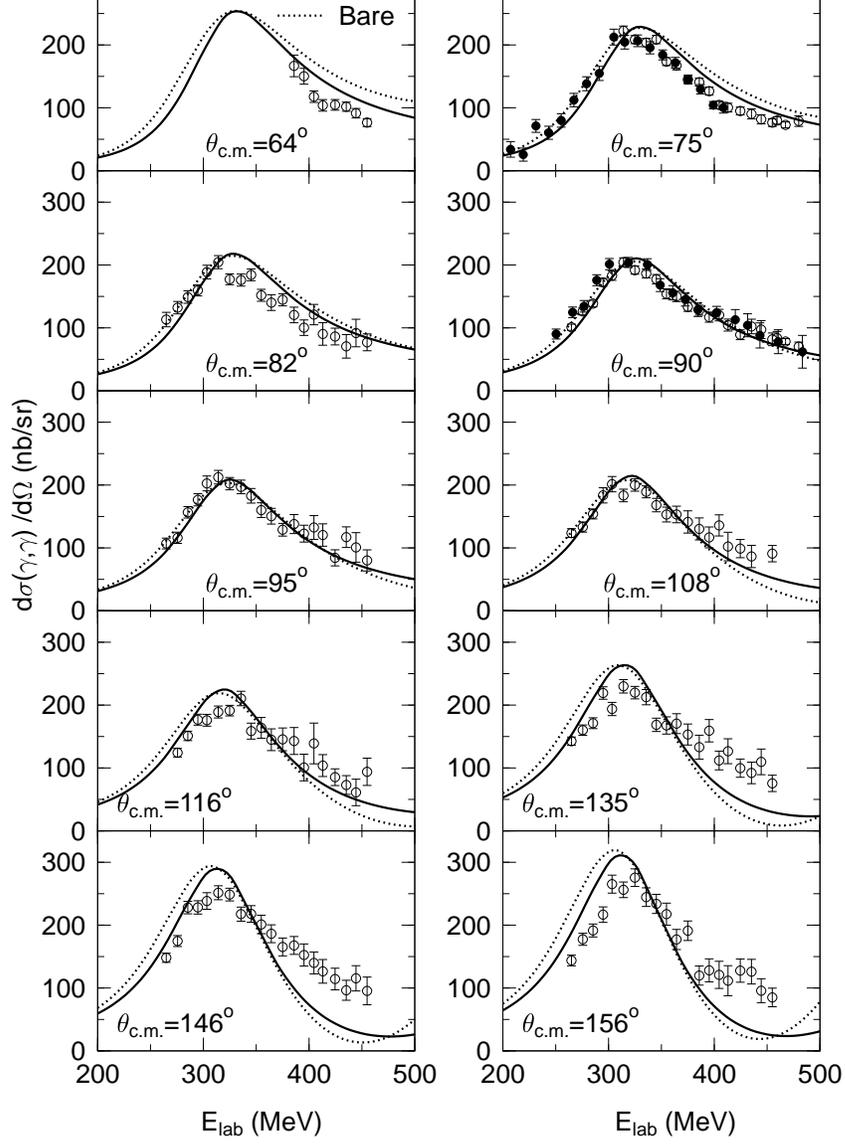}}
\caption[f12p]{
The differential cross section as a functions the photon 
laboratory energy,
for proton Compton scattering at various scattering angles. 
The solid and dotted lines are
from the calculations D and B, respectively.
The data points are taken from \cite{Hun97} (denoted by $\bullet$)
and \cite{Gal01} ($\circ$). 
\figlab{cs_en}}  
\end{figure}

\begin{figure}[!htb]
%\centerline{ \epsfxsize 6cm \epsffile{../figures/fit-b/gam.ps}}
\centerline{ \epsfxsize 11cm \epsffile[30 180 515 610]{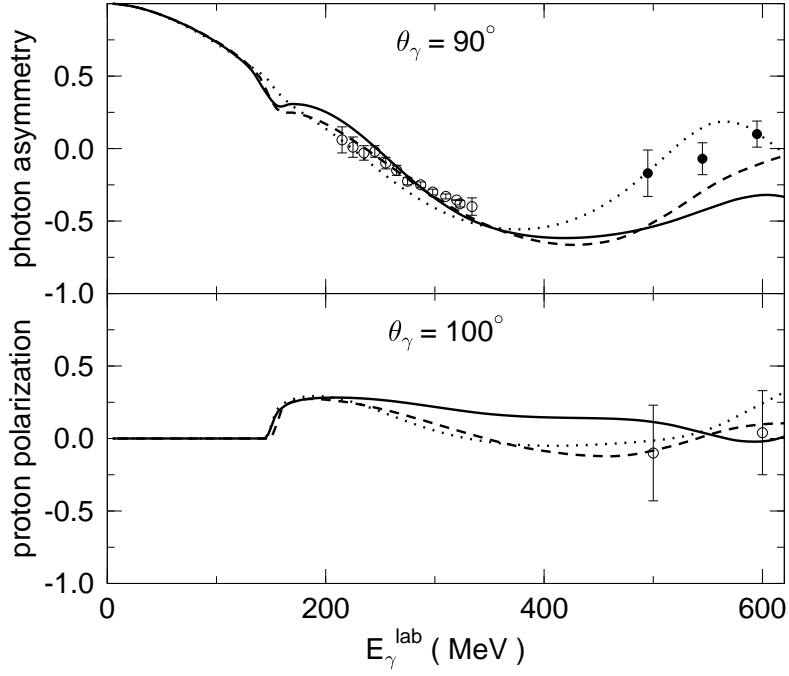}}
\caption[f13]{
%The differential cross section (top),
The photon asymmetry (top)
and proton polarization (bottom) as functions the photon laboratory energy,
for proton Compton scattering. The solid and dotted lines are
from the calculations D and B, respectively.
The results of the dispersion calculation from
Ref.~\cite{Lvo97} are shown by the dashed lines.
The data points are taken from the
following experiments. 
%For the cross section: $\bullet$ \cite{Hun97},
%$\diamond$ \cite{Gen76}, $\star$ \cite{DeW61}, $\circ$ \cite{Tos78};
For the photon asymmetry: $\circ$ \cite{Bla96}, $\bullet$ \cite{Ada93}; 
for the proton polarization: $\circ$ \cite{Wad84}. 
\figlab{gam-qnp}}  
\end{figure}

\newpage
\begin{figure}[!htb]
%\centerline{ \epsfxsize 6cm \epsffile{../figures/fit-b/ee1.ps}}
\centerline{ \epsfxsize 6cm \epsffile{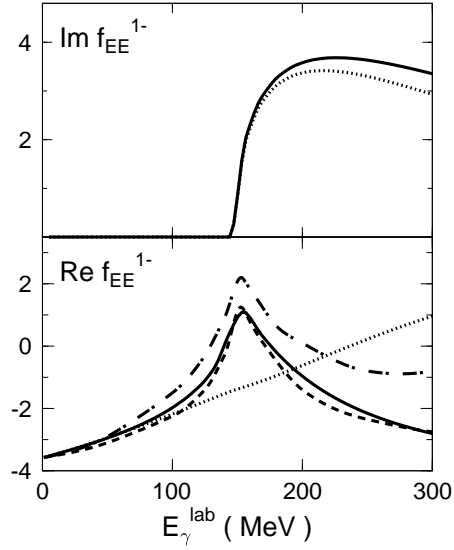}}
\caption[f14]{The
$f_{EE}^{1-}$ partial amplitude of Compton scattering on the
proton in units $10^{- 4}/m_{\pi}$.
Solid line: full calculation D; dotted line: calculation B.
Also shown are the
results of the dispersion analyses of  Ref.~\cite{Pfe74} (dash-dotted line)
and Ref.~\cite{Ber93} (dashed line).
\figlab{fee}} 
\end{figure}

\begin{figure}[!htb]
%\centerline{ \epsfxsize 10.cm \epsffile[0 25 530 110]{diag-g.ps}}
\centerline{ \epsfxsize 10.cm \epsffile[0 25 530 110]{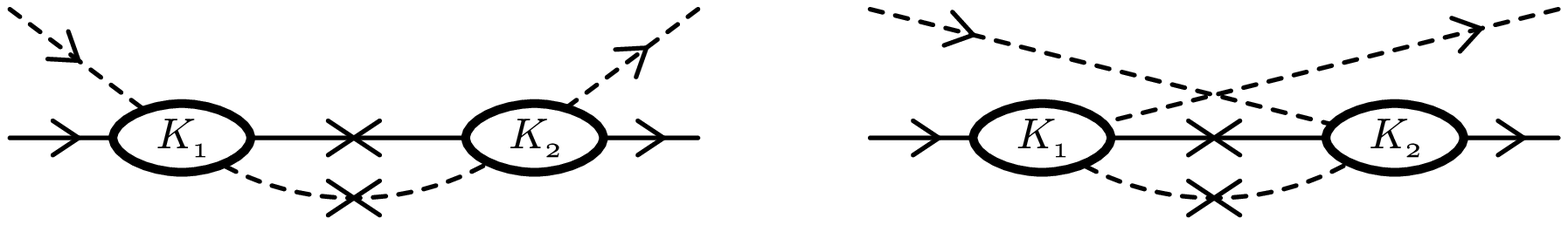}}
\caption[f15]{Left: a diagram contributing to the T-matrix. Pole (on-shell)
contributions from propagators are indicated by crosses. Right: the
crossed version of the diagram on the left. $K_1$ and $K_2$ represent
results of preceding iterations of the kernel.
\figlab{diag-d}}
\end{figure}


\begin{thebibliography}{99}


\bibitem{Low54} F.E. Low, Phys. Rev. {\bf 96}, 1428 (1954);
                M. Gell-Mann and M.L. Goldberger, Phys. Rev. {\bf 96},
                1433 (1954).
\bibitem{Pfe74} W. Pfeil, H. Rollnik, and S. Stankowski, \NPB{73}{1974}{166}.
\bibitem{Ber93} J.C. Bergstrom and E.L. Hallin, \PRC{48}{1993}{1508}.
\bibitem{Lvo97} A.I. L'vov, V.A. Petrun'kin, and M. Schumacher,
                \PRC{55}{1997}{359}.
\bibitem{Hun97} A. H\"{u}nger, J. Peise, A. Robbiano et al.,
                \NPA{62}{1997}{385}.
\bibitem{Bab98} D. Babusci, G. Giordano, A.I. L'vov, G. Matone, and
                A.M. Nathan, \PRC{58}{1998}{1013}.
\bibitem{Dre00}  D. Drechsel, M. Gorchtein, B. Pasquini, and M. Vanderhaeghen,
                 \PRC{61}{2000}{015204}.
\bibitem{Sal51} E.F. Salpeter and H.A. Bethe, \PR{84}{1951}{1232}.
\bibitem{Lah99} A.D. Lahiff and I.R. Afnan \PRC{60}{1999}{024608}.

\bibitem{Pea91} B. C. Pearce and B. K. Jennings, Nucl. Phys. {\bf A528}, 655
                (1991).

\bibitem{Gro93} F. Gross and Y. Surya, Phys. Rev. C {\bf 47}, 703 (1993).

\bibitem{Pas98} V. Pascalutsa and J.A. Tjon, \NPA{631}{1998}{534c};
                   \PLB{435}{98}{245};\PRC{61}{2000}{054003};
                V. Pascalutsa, Ph.D. thesis, University of Utrecht, 1998.
\bibitem{Kon99} S. Kondratyuk and O. Scholten,
                 \PRC{59}{1999}{1070}. % pi-N first
\bibitem{Kon00a} S. Kondratyuk and O. Scholten,
                 \PRC{62}{2000}{025203}.% pi-N extended
\bibitem{Kon01} S. Kondratyuk and O. Scholten,
                \NPA{680}{2001}{175c}. % Adelaide proceedings
\bibitem{Kon00b} S. Kondratyuk and O. Scholten,
                 \NPA{677}{2000}{396}. % gam_N, first
\bibitem{Gou94} P. F. A. Goudsmit, H. J. Leisi, E. Matsinos, B. L. Birbrair,
                and A. B. Gridnev, Nucl. Phys. {\bf A575}, 673 (1994).
\bibitem{Sch96} O. Scholten, A.Yu. Korchin, V. Pascalutsa, and D. Van Neck,
                Phys. Lett. B {\bf 384}, 13 (1996).
\bibitem{Feu98} T. Feuster and U. Mosel, Phys. Rev. C {\bf 58}, 457 (1998).
\bibitem{Kor98} A. Yu. Korchin, O. Scholten, and R.G.E. Timmermans,
                Phys. Lett. B {\bf438}, 1 (1998).
\bibitem{Feu99} T. Feuster and U. Mosel, \PRC{59}{1999}{460}.
\bibitem{Bjo64} J.D. Bjorken and S.D. Drell,
                {\it Relativistic Quantum Mechanics} (McGraw-Hill, 1964);
                C. Itzykson and J.-B. Zuber,
                {\it Quantum Field Theory} (McGraw-Hill, Inc., 1986).
%\bibitem{Kaz59} E. Kazes, Nuovo Cimento {\bf 13}, 1226 (1959).
\bibitem{Man59}  S. Mandelstam, Phys. Rev. {\bf 115}, 1741 (1959);
R. E. Cutkosky, J. Math. Phys. {\bf 1}, 429 (1960); G. 't Hooft and M. J. G.
Veltman, Diagrammar, CERN Yellow Report 73-09;
M. Veltman, Physica 29, 186 (1963).
\bibitem{Bin60} A. Bincer, \PR{118}{1960}{855}.
\bibitem{Gro00} D.E. Groom et. al. , Eur. Phys. J. {\bf C15}, 1 (2000).
\bibitem{Chi61} J.S.R. Chisholm, Nucl. Phys. {\bf 26}, 469 (1961);
                S. Kamefuchi, L. O'Raifeartaigh, and A. Salam,
                Nucl. Phys. {\bf 28}, 529 (1961);
		H.W. Fearing and S. Scherer, \PRC{62}{2000}{034003}.		
\bibitem{Arn95} R.A. Arndt, I.I. Strakovskii, and
                R.L. Workman, Phys. Rev. C {\bf 52}, 2120 (1995).
\bibitem{Arn96} R.A. Arndt, I.I. Strakovskii, and R.L. Workman,
                Phys. Rev. C {\bf 53}, 430 (1996).
\bibitem{Gar93} H. Garcilazo and E. Moya de Guerra, \NPA{562}{1993}{521}.
\bibitem{Hal93} E.L. Hallin {\it et al.}, \PRC{48}{1993}{1497}.
\bibitem{Gal01} G. Galler {\it et al.}, \PLB{503}{2001}{245}.
%\bibitem{Gen76} H. Genzel, M. Jung, R. Wedemeyer, and H.J. Weyer,
%                \ZPA{279}{1976}{399}.
%\bibitem{DeW61} J.W. DeWire, M. Feldman, V.L. Highland, and R. Littauer,
%                \PR{124}{1961}{909}.
%\bibitem{Tos78} K. Toshioka, M. Chiba, S. Kato {\it et al.},
%                \NPB{141}{1978}{364}.
\bibitem{Bla96} G. Blanpeid, M. Blecher, A. Caracappa {\it et al.},
                \PRL{76}{1996}{1023}.
\bibitem{Ada93} F.V. Adamyan, A.Yu. Buniatian, G.S. Frangulian
                {\it et al.}, J. Phys. G: Nucl. Part. Phys. {\bf 19} (1993)
                L139.
\bibitem{Wad84} Y. Wada, K. Egawa, A. Imanishi {\it et al.},
                \NPB{247}{1984}{313}.
\bibitem{Hem98} T.R. Hemmert, B.R. Holstein, and J. Kambor,
                \PRD{57}{1998}{5746}.
\bibitem{Hol00} B.R. Holstein, hep-ph/0010129.
\bibitem{Ber95} V. Bernard, N. Kaiser, and Ulf-G. Meissner, Int. J. Mod.
                Phys.~\VYP{E4}{1995}{193}.
\bibitem{Ber_K93} V. Bernard, N. Kaiser, A. Schmidt, and Ulf-G. Meissner,
                \PLB{319}{1993}{269}.
\bibitem{Gel00} G.C. Gellas, T.R. Hemmert, and Ulf-G. Meissner,
                \PRL{85}{2000}{14}.
\bibitem{Vij00} K.B. Vijaya Kumar, J.A. McGovern, and M.C. Birse,
                \PLB{479}{2000}{167}.
\bibitem{Bir00} M.C. Birse, X. Ji, and J.A. McGovern, nucl-th/0011054.
%\bibitem{Sch91} J. Schmiedmayer, P. Riehs, J.A. Harvey, and N.W. Hill,
%                \PRL{66}{1991}{1015}.
%\bibitem{Koe95} L. Koester {\it et al}, \PRC{51}{1995}{3363}.
\bibitem{Gui78} I. Guiasu, C. Pomponiu, and E.E. Radescu, Ann. Phys. {\bf 114},
                296 (1978).

\end{thebibliography}
\end{document}